\documentclass{article}
\usepackage[top=2cm,bottom=3cm,left=2cm,right=2cm]{geometry}
\usepackage{amsmath}
\usepackage{amsfonts}
\usepackage{stmaryrd}
\usepackage{bm}
\usepackage{hyperref}
\usepackage{upgreek}
\usepackage[mathscr]{euscript}
\usepackage{graphicx}
\usepackage{mathtools}
\usepackage{soul}

\newcommand{\inverttriangle}{%
               \mathrel{\raisebox{.1em}{%
               \reflectbox{\rotatebox[origin=c]{180}{$\triangle$}}}}\!}
               
\numberwithin{equation}{section}
\numberwithin{figure}{section}

\def\eq#1{(\ref{eq:#1})}
\def\lineup{\!\!\!\!\!\!\!\!\!\!&&}

\def\d{\partial}
\def\eps{\epsilon}
\def\Vspace{\phantom{\bigg(}}

\def\deg{\mathrm{deg}}

\def\M{{\bf M}}
\def\m{{\bf m}}
\def\Q{{\bf Q}}
\def\S{{\bf S}}
\def\s{{\bf s}}
\def\T{{\bf T}}

\def\D{{\bf D}}
\def\b{{\bf b}}

\def\P{{\bf P}}
\def\p{{\bf p}}

\def\n{{\bm \upeta}}
\def\mmu{{\bm \upmu}}
\def\ssigma{{\bm \upsigma}}

\def\G{{\bf \hat{G}}}

\def\PsiR{\Psi_\mathrm{R}}
\def\PsiN{\Psi_\mathrm{NS}}

\def\H{\mathcal{H}}
\def\Hr{{\mathcal{H}}^{\mathrm{restricted}}}

\def\wr{\Omega}

\def\[{\big[}
\def\]{\big]}

\begin{document}

\begin{titlepage}
\rightline{\tt LMU-ASC 47/16}
\rightline\today

\begin{center}
\vskip 3.5cm

{\large \bf{Supersymmetry in Open Superstring Field Theory}}

\vskip 1.0cm

{\large {Theodore Erler}}

\vskip 1.0cm

{\it {Arnold Sommerfeld Center, Ludwig-Maximilians University}}\\
{\it {Theresienstrasse 37, 80333 Munich, Germany}}\\
tchovi@gmail.com\\
\vspace{.5cm}

\vskip 2.0cm

{\bf Abstract}

\end{center}

We realize the 16 unbroken supersymmetries on a BPS D-brane as invariances of the action of the corresponding open superstring field theory. We work in the small Hilbert space approach, where a symmetry of the action translates into a symmetry of the associated cyclic $A_\infty$ structure. We compute the supersymmetry algebra, being careful to disentangle the components which produce a translation, a gauge transformation, and a symmetry transformation which vanishes on-shell. Via the minimal model theorem, we illustrate how supersymmetry of the action implies supersymmetry of the tree level open string scattering amplitudes.

\end{titlepage}

\tableofcontents

\section{Introduction}

Following the recent constructions of open superstring field theory \cite{complete,KSR,RWaction}, an important issue to understand is the realization of supersymmetry. Since the string field does not match fermion and boson degrees of freedom off-shell, supersymmetry is not manifest. It is described by a nonlinear transformation of the form
\begin{equation}
\delta_{\mathrm{susy}}\Psi = S_1\Psi + S_2(\Psi,\Psi) + S_3(\Psi,\Psi,\Psi)+ \mathrm{higher\ orders},
\end{equation}
where $S_1,S_2,S_3,...$ are a specific sequence of multi-string products. The goal of this paper is to construct the products of the supersymmetry transformation using the zero mode of the fermion vertex, 
\begin{equation}\oint_{|z|=1} \frac{dz}{2\pi i} \Theta_{a} e^{-\phi/2}(z),\end{equation}
picture changing operators, and Witten's associative string star product. 

We focus on the small Hilbert space formulation of open superstring field theory \cite{KSR,RWaction}, since the supersymmetry transformation in this framework takes a fairly canonical form. In the large Hilbert space formulation \cite{complete} there are ambiguities in the choice of supersymmetry transformation related to the enlarged gauge symmetry of the theory, and we postpone discussion to later work \cite{susyL}.\footnote{Analysis of supersymmetry in the large Hilbert space will appear soon in independent work by H. Kunitomo \cite{KunitomoSUSY}.} Since classical open superstring field theory does not contain gravity, supersymmetry can only be described as a global symmetry. Therefore our analysis is somewhat different in spirit than other recent discussions of supersymmetry in superstring perturbation theory \cite{SenSUSY,revisited}, which utilize the fact that closed superstring field theory incorporates supersymmetry automatically as part of the the local gauge symmetry. Finally, we should emphasize that we only consider unbroken supersymmetries. Describing broken supersymmetries is closely related to the issue of background independence in string field theory, and should be important for understanding the appearance of D-brane charges in the supersymmetry algebra. Further progress in this direction may be possible following~\cite{SenBg,supervac,KOSsing}.

This paper is organized as follows. In section \ref{sec:Ainf} we review the small Hilbert space formulation of open superstring field theory \cite{KSR,RWaction}, mostly to simplify notation and to introduce the concept of {\it cyclic Ramond number} which will be convenient for understanding issues related to cyclicity. In section \ref{sec:smallsusy} we describe the construction of the supersymmetry transformation and prove that it leaves the action invariant. In section \ref{sec:susyalg} we compute the supersymmetry algebra, explicitly describing the gauge transformation and the on-shell trivial symmetry which appear in addition to the momentum operator when computing the commutator of supersymmetry transformations. Finally, in section \ref{sec:min} we use the minimal model to illustrate how supersymmetry of the action implies supersymmetry of the $S$-matrix.

\section{Superstring Field Theory in the Small Hilbert space}
\label{sec:Ainf}

In this section we review the small Hilbert space formulation open superstring field theory, based on an action realizing a cyclic $A_\infty$ structure \cite{KSR,RWaction}. This theory is based on the RNS formulation of the superstring worldsheet, with a $c=15$ matter superconformal field theory tensored with and a $c=-15$ ghost boundary superconformal field theory $b,c,\beta,\gamma$. The $\beta\gamma$ ghosts will be bosonized to the $\xi,\eta,e^\phi$ system \cite{FMS}. The string field is an element of the state space $\H$ of this boundary superconformal field theory. Generally, we consider $\H$ to include both Neveu-Schwarz (NS) and Ramond (R) sector states, as well as states in the small and the large Hilbert space. The small Hilbert space consists of states $A$ satisfying $\eta A=0$, where $\eta$ is the zero mode of the eta ghost, and the large Hilbert space includes states which do not satisfy $\eta A = 0$. So that we can describe fermions and spacetime ghosts, we assume that states in $\H$ can appear in linear combinations with commuting or anticommuting coefficients. In this paper we are interested in supersymmetry, so we require that all states in $\H$ are GSO($+$) projected. 

The following discussion assumes familiarity with the coalgebra representation of $A_\infty$ algebras, in particular as reviewed in \cite{WB}.  In this formalism it is necessary to use a shifted even/odd grading on the open string state space called {\it degree}. The degree of a open string field $A$, denoted $\deg(A)$, is defined to be its Grassmann parity plus~one.

\subsection{Action}

The action can be expressed 
\begin{equation}S = \frac{1}{2}\wr(\Psi,Q\Psi)+\frac{1}{3}\wr(\Psi,M_2(\Psi,\Psi)) + \frac{1}{4}\wr(\Psi,M_3(\Psi,\Psi,\Psi))+\mathrm{higher\ orders}.\label{eq:smallaction}
\end{equation}
There are three main ingredients: A dynamical string field $\Psi$, which includes an NS sector component and a Ramond sector component; a symplectic form $\wr$, which maps two string fields into a number; and multi-string products $M_{n+1}$, which multiply $n+1$ string fields to produce a string field. The 1-string product $M_1$ is equal to the BRST operator $Q$, and the higher products are built from Witten's open string star product with insertions of picture changing operators. Importantly, the products satisfy the relations of a cyclic $A_\infty$ algebra, where the notion of cyclicity is provided by the symplectic form $\wr$. As it happens, the action is purely quadratic in the Ramond string field.  Therefore products $M_{n+1}$ are taken to vanish when multiplying three or more Ramond states.

Let us describe the ingredients of the action in more detail. The dynamical string field $\Psi$ has an NS component and an R component:
\begin{equation}\Psi = \PsiN + \PsiR.\end{equation}
The NS dynamical field $\PsiN$ is a degree even NS state in the small Hilbert space at ghost number 1 and picture $-1$. The Ramond dynamical field $\PsiR$ is a degree even Ramond state in the small Hilbert space at ghost number 1 and picture $-1/2$. In addition, the Ramond string field satisfies the condition \cite{complete}
\begin{equation}\mathscr{X}\mathscr{Y} \PsiR = \PsiR,\label{eq:XY1}\end{equation}
where the operators $\mathscr{X}$ and $\mathscr{Y}$ are defined
\begin{eqnarray}
\mathscr{X} \lineup \equiv  G_0\delta(\beta_0) + b_0\delta'(\beta_0),\\
\mathscr{Y} \lineup \equiv  -c_0\delta'(\gamma_0).
\end{eqnarray}
The operators satisfy
\begin{equation}\mathscr{X}\mathscr{Y}\mathscr{X}= \mathscr{X},\ \ \ \ \mathscr{Y}\mathscr{X}\mathscr{Y} = \mathscr{Y},\ \ \ \ [Q,\mathscr{X}] = 0, \end{equation}
and are BPZ even. Note that $\mathscr{X}$ is singular when acting on states annihilated by $\beta_0$, and $\mathscr{Y}$ is singular when acting on states annihilated by $\gamma_0$. To avoid these singularities, we require that $\mathscr{X}$ only acts on Ramond states in the small Hilbert space at picture $-3/2$, and that $\mathscr{Y}$ only acts on Ramond states in the small Hilbert space at picture $-1/2$ \cite{RWaction}. To describe the subspace of the dynamical string field more efficiently, it is useful to introduce the {\it restricted space}:
\begin{equation}\Hr \subset \H.\end{equation}
The restricted space $\Hr$ consists of NS states in the small Hilbert space at picture $-1$ and Ramond states in the small Hilbert space at picture $-1/2$ which  satisfy the condition $\mathscr{X}\mathscr{Y}A=A$.  The BRST operator preserves this subspace \cite{complete}:
\begin{equation}Q:\Hr\ \rightarrow\ \Hr.\end{equation}
The dynamical string field $\Psi$ is a degree even state in $\Hr$ at ghost number 1.

The symplectic form $\wr$ operates on a pair of states in the restricted space. Accordingly, we will call $\Omega$ the {\it restricted symplectic form}. There are actually three symplectic forms which play an important role:
\begin{eqnarray}
\mathrm{Large\ Hilbert\ space\ symplectic\ form:}\lineup\ \ \ \omega_L(A,B)\ \ \ A,B\in\H, \nonumber \\
\mathrm{Small\ Hilbert\ space\ symplectic\ form:}\lineup\ \ \ \omega_S(A,B)\ \ \ A,B\in \mathrm{small\ Hilbert\ space},\nonumber\\
\mathrm{Restricted\ symplectic\ form:}\lineup\ \ \ \ \wr(A,B)\ \ \ \, A,B\in \Hr.
\end{eqnarray}
All three symplectic forms are graded antisymmetric,
\begin{equation}\omega(A,B) =- (-1)^{\deg(A)\deg(B)}\omega(B,A),\end{equation}
and nondegenerate on their respective domains \cite{complete}. Moreover, the BRST operator satisfies
\begin{equation}
\omega(QA,B) +(-1)^{\deg(A)}\omega(A,QB) = 0\label{eq:Qcyc}
\end{equation}
in all three cases. Generally, an $n$-string product which satisfies
\begin{equation}
\omega(b_n(A_1,...,A_n),A_{n+1})  + (-1)^{\deg(b_n)\deg(A_1)}\omega(A_1,b_n(A_2,...,A_{n+1}))= 0 
\end{equation}
is said to be {\it cyclic} with respect to the symplectic form $\omega$. In particular, the BRST operator is cyclic with respect to all three symplectic forms. The eta zero mode is cyclic with respect to the large Hilbert space symplectic form, and Witten's open string star product
\begin{equation}m_2(A,B) \equiv (-1)^{\deg(A)}A*B\label{eq:m2}\end{equation}
is cyclic with respect to the small and large Hilbert space symplectic forms. Sometimes it will be useful to write the symplectic form as a ``double bra" state $\langle\omega|$, so that 
\begin{equation}\langle\omega|A\otimes B = \omega(A,B).\end{equation}
In this notation, cyclicity of a product $b_n$ can be expressed
\begin{equation}\langle \omega|(b_n\otimes\mathbb{I} + \mathbb{I}\otimes b_n) = 0,\end{equation}
where $\mathbb{I}$ is the identity operator on the state space. The notation can be further simplified using in the coalgebra formalism as 
\begin{equation}
\langle \omega|\pi_2\b_n = 0, \label{eq:cohQcyc}
\end{equation}
where $\pi_2$ is the projector onto the $2$-string component of the tensor algebra and $\b_n$ is the coderivation corresponding to $b_n$. The symplectic forms $\omega_L,\omega_S$ and $\Omega$ mentioned above are defined as follows. The large Hilbert space symplectic form is related to the BPZ inner product in the large Hilbert space by a sign: 
\begin{equation}\omega_L(A,B) \equiv (-1)^{\deg(A)}\langle A,B\rangle_L.\end{equation}
The small Hilbert space symplectic form is be defined by
\begin{equation}\omega_S(a,b)\equiv\omega_L(A,b),\label{eq:wsmall}\end{equation}
where $a,b$ are states in the small Hilbert space and $\eta A = a$. Note that a product $b_n$ which is cyclic with respect to the large Hilbert space symplectic form is also cyclic with respect to the small Hilbert space symplectic form provided $\eta$ acts as a derivation on $b_n$, i.e. $[\n,\b_n]=0$. Finally, the restricted symplectic form is defined by 
\begin{equation}
\Omega(a,b) \equiv \omega_S(\mathcal{G}^{-1}a,b),
\end{equation}
where $a,b$ are states in $\Hr$ and the operator $\mathcal{G}^{-1}$ is defined as
\begin{eqnarray}
\mathcal{G}^{-1}\lineup  = \mathbb{I}\ \ \ \,\mathrm{acting\ on\ NS\ states},\\
\mathcal{G}^{-1}\lineup  = \mathscr{Y}\ \ \ \mathrm{acting\ on\ R\ states}.
\end{eqnarray}
Therefore $\Omega$ is given by the small Hilbert space symplectic form together with an insertion of $\mathscr{Y}$ between pairs of Ramond states.

Finally let us describe the multi-string products $M_{n+1}$. We will call these {\it dynamical products}. The explicit construction will be reviewed in the following subsections, but here we list the essential properties: 
\begin{description}
\item{(M.a)} The dynamical products form an $A_\infty$ algebra. In particular, $M_{n+1}$ is degree odd, and if $\M_{n+1}$ is the coderivation corresponding to $M_{n+1}$, the sum
\begin{equation}\M = \sum_{n=0}^\infty \M_{n+1}\end{equation}
is a nilpotent coderivation in the tensor algebra:\footnote{The commutator $[,]$ is graded with respect to degree.}
\begin{equation}\[\M,\M\] = 0.\end{equation}
\item{(M.b)} The dynamical products carry the appropriate ghost number and picture so that the equations of motion,
\begin{equation}0=Q\Psi + M_2(\Psi,\Psi) + M_3(\Psi,\Psi,\Psi) + \mathrm{higher\ orders},\end{equation}
carry ghost number 2 and picture $-1$ in the NS sector, and ghost number 2 and picture $-1/2$ in the Ramond sector.
\item{(M.c)} The eta zero mode acts as a derivation of the dynamical products. Equivalently,
\begin{equation}\[\n,\M\] = 0,\end{equation}
where $\n$ is the coderivation corresponding to $\eta$. This implies that the dynamical products multiply consistently inside the small Hilbert space.
\item{(M.d)} The dynamical products preserve the Ramond constraint $\mathscr{X}\mathscr{Y} = 1$ when acting on states in $\Hr$.
\item{(M.e)} The dynamical products are cyclic with 
respect to the restricted symplectic form:
\begin{equation}\langle \wr|\pi_2\M = 0\ \ \ \ \mathrm{on}\ \ T\Hr.\end{equation}
\end{description}
We can summarize these conditions as requiring that the dynamical products $M_{n+1}$ define a cyclic $A_\infty$ algebra on $\Hr$. In particular, conditions (M.b), (M.c) and (M.d) imply that the restricted space is closed under multiplication with $M_{n+1}$. Note that conditions (M.a), (M.b) and (M.c) are sufficient for constructing gauge invariant equations of motion, as described in \cite{Ramond}. We additionally require conditions (M.d) and (M.e) to have a gauge invariant action.

\subsection{Counting Ramond States}
\label{subsec:count}

To construct the dynamical products it is necessary to introduce some notation for keeping track of the number of Ramond states that are multiplied with a product. We start by considering the tensor algebra generated from the open string state space $\H$:
\begin{equation}
T\H = \H^{\otimes 0}\,\oplus\,\H\,\oplus\,\H^{\otimes 2}\,\oplus\,\H^{\otimes 3}\oplus...\ .
\end{equation}
We introduce a projection operator $\pi_m$, 
\begin{equation}
\pi_m:T\H \to T\H,\ \ \ \ \pi_m\pi_n = \delta_{mn}\pi_m,
\end{equation}
which projects onto the $m$-string subspace of the tensor algebra. We also consider a projection operator $\pi^r$,
\begin{equation}
\pi^r:T\H \to T\H,\ \ \ \ \pi^r\pi^s = \delta^{rs}\pi^r,
\end{equation}
which selects multi-string states in the tensor algebra containing $r$ Ramond factors (but an undetermined number of NS factors). Multiplying $\pi_m$ and $\pi^r$ defines a projection operator
\begin{eqnarray}
\pi_m^r \equiv \pi_m\pi^r =\pi^r\pi_m,
\end{eqnarray}
which selects $m$-string states containing $r$ Ramond factors. We introduce a coderivation ${\bf 1}$ which satisfies
\begin{equation}{\bf 1}\pi_m = m \pi_m.\end{equation}
The eigenvalue of ${\bf 1}$ counts the total number of states. We also introduce a coderivation ${\bf R}$ which satisfies
\begin{equation}{\bf R}\pi^r = r\pi^r.\end{equation}
The eigenvalue of ${\bf R}$ counts the number of Ramond states.  

Consider an operator on the tensor algebra $\bm{\mathcal{O}}_{n+1}$ which has well defined eigenvalue under commutation with ${\bf 1}$:
\begin{equation}\[\bm{\mathcal{O}}_{n+1},{\bf 1}\] = n\bm{\mathcal{O}}_{n+1}.\end{equation}
The subscript $n+1$ denotes the integer eigenvalue $n$. The operator commutes through the projector $\pi_m$ as
\begin{equation}
\pi_m\bm{\mathcal{O}}_{n+1} = \bm{\mathcal{O}}_{n+1}\pi_{m+n},
\end{equation}
which means that $\bm{\mathcal{O}}_{n+1}$ removes $n$ states from the tensor algebra. Next consider an operator $\bm{\mathcal{O}}|_r$ which has well defined eigenvalue under commutation with ${\bf R}$
\begin{equation}\[\bm{\mathcal{O}}|_r,{\bf R}\] = r\bm{\mathcal{O}}|_r.\label{eq:OReig}\end{equation}
We refer to the integer eigenvalue $r$ as the {\it Ramond number} of $\bm{\mathcal{O}}|_r$. The Ramond number of an operator on the tensor algebra will be indicated by a vertical slash followed by a subscript. Such an operator commutes through the projector $\pi^r$ as
\begin{equation}
\pi^s\bm{\mathcal{O}}|_{r} = \bm{\mathcal{O}}|_{r}\pi^{r+s},\label{eq:pircom}
\end{equation}
which means that $\bm{\mathcal{O}}|_r$ removes $r$ Ramond states from the tensor algebra.

An important case is when the operators are coderivations. A coderivation $\b_{n+1}$ which satisfies 
\begin{equation}\[\b_{n+1},{\bf 1}\] = n \b_{n+1}, \label{eq:1eig}\end{equation}
is characterized by a corresponding $(n+1)$-product $b_{n+1}$:
\begin{equation}\pi_1\b_{n+1} = b_{n+1}\pi_{n+1}.\end{equation}
The coderivation property uniquely determines a coderivation $\b_{n+1}$ once the product $b_{n+1}$ has been defined (see e.g. \cite{WB}). We may also consider a coderivation $\b|_r$ which carries definite Ramond number: 
\begin{equation}\[\b|_r,{\bf R}\] = r\b|_r.\label{eq:Reig}\end{equation}
Since ${\bf  R}$ and ${\bf 1}$ commute, we can have a coderivation $\b_{n+1}|_r$ which simultaneously has well defined eigenvalue under commutation with ${\bf 1}$ and ${\bf R}$. Such a coderivation satisfies 
\begin{equation}\pi^s_m\b_{n+1}|_r = \b_{n+1}|_r\pi_{m+n}^{r+s},\end{equation}
and is uniquely defined by an $(n+1)$-string product, which we write $b_{n+1}|_r$. We say that the product $b_{n+1}|_r$ carries {\it Ramond number} $r$. A product of Ramond number $r$ can be nonzero only when the number of Ramond inputs minus the number of Ramond outputs is equal to $r$. This means that $b_{n+1}|_r$ must satisfy
\begin{eqnarray}
b_{n+1}|_r\Big(r\ \mathrm{Ramond\ states}\Big) \lineup = \mathrm{NS\ state},\nonumber\\
b_{n+1}|_r\Big(r+1\ \mathrm{Ramond\ states}\Big) \lineup = \mathrm{R\ state},\nonumber\\
b_{n+1}|_r\Big(\mathrm{otherwise}\Big) \lineup = 0.
\end{eqnarray}
A generic product $b_m$ can be written as a sum of products of definite Ramond number
\begin{equation}
b_{n+1} = b_{n+1}|_{-1} + b_{n+1}|_0 + b_{n+1}|_1+...+b_{n+1}|_{n+1}.
\end{equation}
The Ramond number is bounded between $-1$ and $m$ since $b_m$ can have at most $m$ Ramond inputs and $1$ Ramond output. The BRST operator carries Ramond number zero
\begin{equation}Q= Q|_0.\end{equation}
Note that Ramond number adds when composing products. Therefore Ramond number defines a grading on the space of products and coderivations, which is of central importance in obtaining a solution of $A_\infty$ relations.

However, the concept of Ramond number is less useful when it comes to questions of cyclicity. To see why, let $\omega\circ b_{n+1}$ denote the cyclic permutation of a product $b_{n+1}$, defined through the relation \cite{WB}
\begin{equation}\langle \omega_L|\mathbb{I}\otimes b_{n+1} = -\langle\omega_L|(\omega\circ b_{n+1}) \otimes  \mathbb{I}.
\end{equation}
If $b_{n+1}|_r$ carries Ramond number $r$, generally $\omega\circ (b_{n+1}|_r)$ cannot have definite Ramond number (except, in some cases, when the Ramond number is zero.). This means that products of definite Ramond number are not cyclic. Therefore it is useful to introduce a different notion of Ramond number which is invariant under cyclic permutations of products. We call this {\it cyclic Ramond number}. The product $b_{n+1}|^r$ has cyclic Ramond number $r$ if the number of Ramond inputs {\it plus} the number of Ramond outputs is equal to $r$. Thus we have 
\begin{eqnarray}
b_{n+1}|^r\Big(r\ \mathrm{Ramond\ states}\Big) \lineup = \mathrm{NS\ state},\nonumber\\
b_{n+1}|^r\Big(r-1\ \mathrm{Ramond\ states}\Big) \lineup = \mathrm{R\ state},\nonumber\\
b_{n+1}|^r\Big(\mathrm{otherwise}\Big) \lineup = 0.
\end{eqnarray}
Cyclic Ramond number is denoted with a vertical slash followed by a superscript. It is clear from this definition that $\omega\circ(b_{n+1}|^r)$ has the same cyclic Ramond number as $b_{n+1}|^r$. However, cyclic Ramond number does not add when composing products, and is less useful for the analysis of  $A_\infty$ relations. 

We may consider products which simultaneously have definite Ramond number $r$ and cyclic Ramond number~$s$. We write such products as $b_{n+1}|_r^s$. A product $b_{n+1}|_r^s$ can only be nonzero only if $s=r$ or if $s=r+2$. In these cases, we have
\begin{eqnarray}
b_{n+1}|_r^r\Big(r\ \mathrm{Ramond\ states}\Big) \lineup = \mathrm{NS\ state},\nonumber\\
b_{n+1}|_r^r\Big(\mathrm{otherwise}\Big) \lineup = 0,
\end{eqnarray}
and 
\begin{eqnarray}
b_{n+1}|_r^{r+2}\Big(r+1\ \mathrm{Ramond\ states}\Big) \lineup = \mathrm{R\ state},\nonumber\\
b_{n+1}|_r^{r+2}\Big(\mathrm{otherwise}\Big) \lineup = 0.
\end{eqnarray}
Sometimes it is useful to view the ``vertical slash" as an operation which selects the component of a product with the indicated Ramond number and/or cyclic Ramond number. Thus if we are given a product $b_{n+1}$, we can apply the operation $|_r^s$ to arrive at the product $b_{n+1}|_r^s$. This operation may be defined in terms of the projection operators~$\pi_m^r$:
\begin{eqnarray}
\Big(b_{n+1}|_r^r\Big)\pi_{n+1} \lineup \equiv \pi_1^0\Big(\b_{n+1}\Big)\pi_{n+1}^r,\label{eq:slashrr}\\
\Big(b_{n+1}|_r^{r+2}\Big)\pi_{n+1} \lineup \equiv \pi_1^1\Big(\b_{n+1}\Big)\pi_{n+1}^{r+1}.\label{eq:slashrrp2}
\end{eqnarray}
We also define
\begin{eqnarray}
|_r \lineup \equiv |_r^r + |_r^{r+2},\\
|^r \lineup \equiv |_{r-2}^r + |_r^r.\label{eq:cycr}
\end{eqnarray}
The action of $|_r^s$ on products naturally defines an action of $|_r^s$ on coderivations. Note that $b_{n+1}|_r^s$ does not always derive from operating $|_r^s$ on a product $b_{n+1}$ defined for generic Ramond numbers. When this is the case, it should be clear from context.

\subsection{Dynamical Products}
\label{subsec:dynamical}

The dynamical products $M_{n+2}$ are built by taking compositions of Witten's open string star product with insertions of picture changing operators. 
The BRST operator, the eta zero mode, and the star product define three mutually commuting $A_\infty$ structures:
\begin{eqnarray}
\lineup \[\Q,\Q\]=0,\ \ \ \ \, \[\n,\n\]= 0,\ \ \ \ \ \[\Q,\n\]=0,\nonumber\\
\lineup \[\Q,\m_2\]=0, \ \ \ \[\n,\m_2\]=0,\ \ \ \[\m_2,\m_2\]=0.\label{eq:3Ainf}
\end{eqnarray}
These relations say that $Q$ and $\eta$ are nilpotent and anticommute, are derivations of the open string star product, and the open string star product is associative. We can expand these relations further by taking components of definite Ramond number. The open string star product has a component at Ramond number $0$ and at Ramond number $2$:
\begin{equation}m_2 = m_2|_0 + m_2|_2\end{equation}
Taking the Ramond number $0$ and $2$ components of \eq{3Ainf} implies
\begin{eqnarray}
\[\Q,\m_2|_0\]\lineup = 0,\ \ \ \ \ \ \, \[\Q,\m_2|_2\] = 0,\label{eq:QRder}\\
\ \[\n,\m_2|_0\]\lineup = 0,\ \ \ \ \ \ \ \, \[\n,\m_2|_2\] = 0,\\
\ \[\m_2|_0,\m_2|_0\]\lineup = 0,\ \ \ \[\m_2|_0,\m_2|_2\] = 0.\label{eq:Rass}
\end{eqnarray}
The commutator $\[\m_2|_2,\m_2|_2\]$ automatically vanishes since a 3-string product cannot carry Ramond number $4$.  

The dynamical products have components at Ramond number zero and two:
\begin{equation}M_{n+2} = M_{n+2}|_0 + m_{n+2}|_2.\label{eq:MtR}\end{equation}
In particular, $M_{n+1}$ vanishes when multiplying four or more Ramond states. In fact, $M_{n+1}$ will also vanish when multiplying three Ramond states, so the action is quadratic in the Ramond string field. The product $M_1|_0$ is identified with the BRST operator $Q$, and the product $m_2|_2$ is identified with the Ramond number $2$ component of Witten's open string star product. To construct the dynamical products we introduce auxiliary multi-string products:
\begin{eqnarray}
\mathrm{bare\ products:}\ \ \ m_{n+2}|_0 \lineup\ \ \  \mathrm{degree\ odd},\nonumber\\
\mathrm{gauge\ products:}\ \ \ \mu_{n+2}|_0 \lineup \ \ \ \mathrm{degree\ even}.
\end{eqnarray}
The bare 2-product $m_2|_0$ is the Ramond number zero component of Witten's open string star product. We promote $m_{n+2}|_0$ and $\mu_{n+2}|_0$ to coderivations $\m_{n+2}|_0$ and $\mmu_{n+2}|_0$, and define generating functions
\begin{eqnarray}
\m|_0(t) \lineup \equiv \sum_{n=0}^\infty t^n\m_{n+2}|_0,\\
\mmu|_0(t) \lineup\equiv\sum_{n=0}^\infty t^n\mmu_{n+2}|_0,
\end{eqnarray}
satisfying the equations
\begin{eqnarray}
\frac{d}{dt}\m|_0(t) \lineup = \[\m|_0(t),\mmu|_0(t)\],\label{eq:mdiff}\\
\ \[\n,\mmu|_0(t)\] \lineup = \m|_0(t).\Vspace\label{eq:mudiff}
\end{eqnarray}
Expanding in powers of $t$, this turns into a recursive system of equations determining the higher order gauge products and bare products in terms of lower order ones: 
\begin{eqnarray}
\m_{n+3}|_0\lineup = \frac{1}{n+1}\sum_{k=0}^n \[\m_{k+2}|_0,\mmu_{n-k+2}|_0\],\label{eq:mrec}\\
\ \[\n,\mmu_{n+2}|_0\]\lineup = \m_{n+2}|_0.\label{eq:murec}
\end{eqnarray}
The last equation should be solved to determine $\mu_{n+2}|_0$ in terms of $m_{n+2}|_0$. This requires a choice of contracting homotopy for $\n$, which determines a configuration of picture changing insertions in the vertices. We will explain how to solve \eq{murec} in a moment. Once we have solved \eq{mrec} and \eq{murec}, we construct the dynamical products as follows. Define generating functions
\begin{eqnarray}
\M|_0(t) \lineup \equiv \sum_{n=0}^\infty t^n\M_{n+1}|_0,\\
\m|_2(t) \lineup\equiv\sum_{n=0}^\infty t^n\m_{n+2}|_2,
\end{eqnarray}
satisfying the differential equations
\begin{eqnarray}
\frac{d}{dt}\M|_0(t) \lineup = \[\M|_0(t),\mmu|_0(t)\], \label{eq:Mdiff}\\
\frac{d}{dt}\m|_2(t) \lineup = \[\m|_2(t),\mmu|_0(t)\].\label{eq:m2diff}
\end{eqnarray}
Expanding in powers of $t$, this turns into a recursive system of equations: 
\begin{eqnarray}
\M_{n+2}|_0\lineup = \frac{1}{n+1}\sum_{k=0}^n \[\M_{k+1}|_0,\mmu_{n-k+2}|_0\],\\
\m_{n+3}|_2\lineup = \frac{1}{n+1}\sum_{k=0}^n \[\m_{k+2}|_2,\mmu_{n-k+2}|_0\].
\end{eqnarray}
Solving this recursion gives the dynamical products as
\begin{equation}\M = \M|_0+\m|_2.\end{equation}
Here and in what follows all generating functions are evaluated at $t=1$ when the dependence on $t$ is not explicitly indicated.

The construction so far gives dynamical products satisfying conditions (M.a), (M.b) and (M.c) \cite{Ramond}. Implementing conditions (M.d) and (M.e), however, requires a specific choice of contracting homotopy for $\n$ when defining the gauge products from \eq{murec}. The contracting homotopy chosen in \cite{RWaction} uses a picture changing insertion:\footnote{In \cite{RWaction} $\Xi$ was denoted $\widetilde{\xi}$.} 
\begin{equation}
\Xi:\ \mathrm{degree\ odd},\ \ \mathrm{ghost\ number}\ -1,\ \ \mathrm{picture}\ 1,
\end{equation}
and we also define
\begin{equation}X \equiv [Q,\Xi].\end{equation}
$\Xi$ has the following properties: 
\begin{description}
\item{1)} $\Xi$ is a contracting homotopy for $\eta$: $[\eta,\Xi] = \mathbb{I}$.
\item{2)} $\Xi$ is BPZ even: $\langle \omega_L|\Xi\otimes\mathbb{I} = \langle \omega_L|\mathbb{I}\otimes\Xi$.
\item{3)} $\Xi$ and $X$ are defined acting on generic states in $\H$ (unlike, in particular, the operator $\mathscr{X}$).
\item{4)} $X = \mathscr{X}$ when acting on a Ramond state at picture $-3/2$ in the small Hilbert space.
\item{5)} $\Xi$ is nilpotent: $\Xi^2 = 0$.\footnote{The relation $\Xi^2=0$ is not needed for the dynamical products, but it will be needed for the supersymmetry transformation.}
\end{description}
The definition of $\Xi$ is reviewed in appendix \ref{app:operator}. We then define the gauge products according to \cite{KSR,RWaction}
\begin{eqnarray}
\mu_{n+2}|_0^0 \lineup \equiv \frac{1}{n+3}\Big(\Xi m_{n+2}|_0^0 + m_{n+2}|_0^0(\Xi\otimes\mathbb{I}^{\otimes n+1}+ ...+\mathbb{I}^{\otimes n+1}\otimes\Xi)\Big),\label{eq:mu00}\\
\mu_{n+2}|_0^2 \lineup \equiv \Xi m_{n+2}|_0^2.\label{eq:mu02}
\end{eqnarray}
Note that $\mu_{n+2}|_0$ has components at cyclic Ramond number $0$ and $2$, and these components must be chosen differently. This definition implies that the dynamical products satisfy \cite{RWaction}: 
\begin{equation}M_{n+2}|_0^2 = X m_{n+2}|_0^2,\ \ \ \ m_{n+2}|_2^4 = 0.\label{eq:specprod}\end{equation}
The first relation implies that the dynamical products are consistent with the constraint $\mathscr{X}\mathscr{Y}A=A$ in the Ramond sector, as required by condition (M.d), and the second relation implies that the dynamical products vanish when multiplying three or more string fields. It is useful to express the dynamical products in components of definite cyclic Ramond number. Using \eq{specprod} we have
\begin{eqnarray}M_{n+2} \lineup = M_{n+2}|_0+m_{n+2}|_2\nonumber\\
\lineup = M_{n+2}|_0^0 + M_{n+2}|_0^2 + m_{n+2}|_2^2 + m_{n+2}|_2^4\nonumber\\
\lineup = M_{n+2}|_0^0 + X m_{n+2}|_0^2 + m_{n+2}|_2^2.\label{eq:MRcR}\end{eqnarray}
To further simplify, it is useful to combine $m_{n+2}|_0$ with $m_{n+2}|_2$ into a single product
\begin{eqnarray}
m_{n+2}\lineup \equiv m_{n+2}|_0 + m_{n+2}|_2
\end{eqnarray}
and introduce the operator \cite{1PIR}
\begin{equation}\mathcal{G} \equiv \mathbb{I}|^0 + X|^2,\label{eq:SenG}\end{equation}
which acts as the identity on NS states and as $X$ on Ramond states. Then \eq{MRcR} can be expressed  
\begin{equation}M_{n+2} = \mathcal{G}(M_{n+2}|^0 + m_{n+2}|^2).\label{eq:cycMt0}\end{equation}
Therefore the dynamical products have a component at cyclic Ramond number $0$ and a component at cyclic Ramond number $2$. Using coderivations we may write this as\footnote{Given an operator  $\mathcal{O}:\H\to\H$ and a coderivation $\b$ defined by multi-string products $b_n$, we use $\mathcal{O}\b$ to denote the coderivation defined by the products $\mathcal{O}b_n$. }.
\begin{equation}\M = \mathcal{G}(\M|^0 + \m|^2),\label{eq:cycMt}\end{equation}
where $\m\equiv \m|_0+\m|_2$.  To appreciate the structure of \eq{cycMt}, recall the restricted symplectic form contains the operator $\mathcal{G}^{-1}$:
\begin{equation}
\mathcal{G}^{-1} = \mathbb{I}|^0 + \mathscr{Y}|^2.
\end{equation}
We have the relation
\begin{equation}\mathcal{G}\mathcal{G}^{-1} = \mathbb{I}\ \ \ \mathrm{on}\ \ \Hr,\end{equation}
since $\mathscr{X}\mathscr{Y}$ acts as the identity on the restricted space. From this it is clear that the factor of $\mathcal{G}$ in \eq{cycMt} is required to cancel the factor of $\mathcal{G}^{-1}$ in the restricted symplectic form. Then cyclicity of $M_{n+2}$ translates to the statement that $M_{n+2}|^0$ and $m_{n+2}|^2$ are cyclic with respect to the small Hilbert space symplectic form.  

It is useful to recall that the construction of $\M$ is equivalent to the construction of a field redefinition which relates $\M$ to comparatively simple $A_\infty$ structure \cite{WB,Ramond}. This can be understood by introducing the cohomomorphism 
\begin{equation}
\G \equiv \mathcal{P}\exp\left[\int_0^1 ds\,\mmu|_0(s)\right],\label{eq:G}
\end{equation}
where the path ordering is from left to right in sequence of increasing $s$. We also define $\G(t)$ by replacing the upper limit of the integral in the path ordered exponential with $t$. From this cohomomorphism the gauge products can be computed using
\begin{equation}
\mmu|_0(t) = \G(t)^{-1}\frac{d}{dt}\G(t).
\end{equation}
Moreover, any coderivation $\b(t)$ which satisfies the differential equation
\begin{equation}
\frac{d}{dt}\b(t)=\[\b(t),\mmu|_0(t)\] 
\end{equation}
can be expressed
\begin{equation}
\b(t) = \G(t)^{-1}\b(0)\G(t).
\end{equation}
This implies the formulas
\begin{eqnarray}
\M|_0 \lineup = \G^{-1}\Q\G,\\
\m|_0 \lineup = \G^{-1}\m_2|_0\G,\\
\m|_2 \lineup = \G^{-1}\m_2|_2\G,
\end{eqnarray}
which, together with \eq{mudiff}, imply \cite{OkWB,Ramond}
\begin{eqnarray}
\M \lineup = \G^{-1}(\Q+\m_2|_2)\G,\Vspace\label{eq:GMt}\\
\n \lineup = \G^{-1}(\n - \m_2|_0)\G.\Vspace\label{eq:Gn}
\end{eqnarray}
Therefore, $\M$ can be derived by a similarity transformation from two comparatively simple $A_\infty$ structures:
\begin{equation}
\Q+\m_2|_2,\ \ \ \ \n - \m_2|_0.
\end{equation}
Consider the field redefinition 
\begin{equation}
\frac{1}{1-\varphi} = \pi_1\G\frac{1}{1-\Psi},\label{eq:improp}
\end{equation}
where $\varphi$ is a new dynamical string field and
\begin{equation}
\frac{1}{1-A} \equiv 1_{T\H} \ +\  A\  +\, A\otimes A + A\otimes A\otimes A + ...
\end{equation}
denotes the group-like element of a degree even string field $A$. In \cite{WB}, the transformation from $\Psi$ to $\varphi$ was called an {\it improper} field redefinition, since it does not preserve the small Hilbert space constraint on the string field. The equations of motion and small Hilbert space constraint of $\Psi$ 
\begin{eqnarray}
0\lineup = \M\frac{1}{1-\Psi},\label{eq:varphiEOM}\\
0\lineup = \n\frac{1}{1-\Psi},\label{eq:varphiconst}
\end{eqnarray}
transform into 
\begin{eqnarray}
0\lineup = (\Q+\m_2|_2)\frac{1}{1-\varphi},\\
0\lineup = (\n - \m_2|_0)\frac{1}{1-\varphi}.
\end{eqnarray}
Projecting on to the 1-string component gives Chern-Simons-like equations \cite{BerkRamond}
\begin{equation}
(Q-\eta)\varphi + \varphi*\varphi = 0.\label{eq:CSEOM}
\end{equation}
The $A_\infty$ superstring field theory can be viewed as one approach to deriving these equations from an action. 

\section{Supersymmetry Transformation}
\label{sec:smallsusy}

We are now ready to discuss supersymmetry. We consider open superstring field theory formulated on a maximally supersymmetric D-brane. The goal is to find a transformation of the dynamical string field $\Psi$ realizing all sixteen unbroken supersymmetries. The natural place to start \cite{FMS} is the zero-mode of the fermion vertex in the $-1/2$ picture:
\begin{equation}s_1 \equiv \sqrt{2}\oint_{|z|=1} \frac{dz}{2\pi i} \Theta_{a} e^{-\phi/2}(z) \eps_{a}.\label{eq:s1}\end{equation}
Let us explain the notation. The index on $s_1$ indicates that this operator is a 1-string product. The $\sqrt{2}$ factor is included to obtain the canonical normalization of the supersymmetry algebra. The operator $\Theta_{a}$ is the spin field:
\begin{equation}
\Theta_{a}(z) \equiv \exp\left[\sum_{j=0}^4 a_j H_j\right](z),\ \ \ a = (a_0,a_1,a_2,a_3,a_4),\ \ \ a_j=\pm\frac{1}{2},
\end{equation}
where the scalars $H_i$ realize the bosonization of the worldsheet fermions $\psi^\mu$ through\footnote{Our conventions for bosonization, spinors, and gamma matrices follows \cite{Polchinski}.}
\begin{equation}
\frac{1}{\sqrt{2}}(\psi^0+\psi^1)=e^{i H_0},\ \ \ \ \ \frac{1}{\sqrt{2}}(\psi^{2j}+i\psi^{2j+1})=e^{i H_j}\ \ j=1,...,4.
\end{equation}
The object $\eps_{a}$ is a supersymmetry parameter---a constant degree odd spinor. The repeated spinor index $a$ is summed. To keep notation simple we leave the dependence of $s_1$ on the supersymmetry parameter implicit. Since we make a GSO($+$) projection in both NS and R sectors, the supersymmetry parameter must have positive chirality. Therefore $s_1$ may represent 16 independent supersymmetries. The massless fermions on the D-brane are described by the vertex operator $c\Theta_{a}e^{-\phi/2}$ multiplied by an anticommuting spinor field. Since this should describe a degree even state,\footnote{Note that the degree of a vertex operator is opposite to the degree of the associated string field.} the operator $\Theta_{a}e^{-\phi/2}$ must be degree odd for positive chirality $a$. Therefore $s_1$ is degree even, and carries ghost number $0$ and picture $-1/2$.

The operator $s_1$ has the following algebraic properties:  
\begin{equation}\[\Q,\s_1\] = 0,\ \ \ \[\n,\s_1\] = 0,\ \ \ \[\s_1,\m_2\] = 0.\ \ \ \end{equation}
In particular, $s_1$ commutes with $Q$ and $\eta$, and, since it is the zero mode of a weight one primary, is a derivation of the open string star product. Also $s_1$ is BPZ odd,
\begin{equation}\omega_L(A,s_1B) = -\omega_L(s_1A,B), \label{eq:s1cyc}\end{equation}
and is therefore cyclic with respect to the large Hilbert space (and small Hilbert space) symplectic form. 

\subsection{Supersymmetry in the Free Theory}
\label{subsec:freesusy}

Let's start by considering the supersymmetry transformation in the free theory:
\begin{equation}S_\mathrm{free} = \frac{1}{2}\wr(\Psi,Q\Psi).\end{equation}
We assume that the NS string field transforms as 
\begin{equation}
\delta_{\mathrm{susy}} \PsiN = s_1\PsiR.\label{eq:linsusyNS}
\end{equation}
The Ramond string field cannot transform as $s_1\PsiN$, since this carries the wrong picture and is inconsistent with the constraint $\mathscr{X}\mathscr{Y}A=A$. These problems can be solved simultaneously by multiplying the transformation by $\mathscr{X}$:
\begin{equation}
\delta_{\mathrm{susy}} \PsiR = \mathscr{X}s_1\PsiN.\label{eq:linsusyR}
\end{equation}
The pictures match up on both sides, and the constraint is satisfied due to $\mathscr{X}\mathscr{Y}\mathscr{X} = \mathscr{X}$. We can package NS and R supersymmetry transformations together in the form
\begin{equation}
\delta_{\mathrm{susy}}\Psi = S_1\Psi,\label{eq:linsusy}
\end{equation}
for the appropriately defined operator $S_1$. The operator $S_1$ can be decomposed into a piece at Ramond number $-1$ and a piece at Ramond number $1$:
\begin{equation}S_1 = X s_1|_{-1} + s_1|_1.\label{eq:tS1}\end{equation}
It is convenient to replace $\mathscr{X}$ with $X$ so that $S_1$ is defined acting on arbitrary states in $\H$. We can equivalently express $S_1$ as an operator of cyclic Ramond number $1$:
\begin{equation}
S_1 = \mathcal{G} s_1|^1,
\end{equation}
using $\mathcal{G}$ from \eq{SenG}. Here the superscript for cyclic Ramond number is redundant since $s_1|^1 = s_1$.

Let us demonstrate that this is a symmetry of the free action. This relies on two properties:
\begin{eqnarray}
[Q,S_1\] = 0,\ \ \ \ \ 
\wr(S_1A,B) + \wr(A,S_1B) = 0.
\end{eqnarray}
It is easy to see that $S_1$ is BRST invariant because both $X$ and $s_1$ are BRST invariant. The fact that $S_1$ is cyclic with respect to $\wr$ can be shown as follows:
\begin{eqnarray}
\langle \wr|(S_1\otimes\mathbb{I} +\mathbb{I}\otimes S_1)\lineup = \langle\omega_S| (\mathcal{G}^{-1}\otimes\mathbb{I})(\mathcal{G}s_1\otimes \mathbb{I}+\mathbb{I}\otimes \mathcal{G}s_1)\nonumber\\
\lineup = \langle\omega_S| (s_1\otimes\mathcal{G}\mathcal{G}^{-1}+\mathcal{G}\mathcal{G}^{-1}\otimes s_1)\nonumber\\
\lineup = \langle\omega_S| (s_1\otimes\mathbb{I}+\mathbb{I}\otimes s_1)\nonumber\\
\lineup = 0,\label{eq:Gcan}
\end{eqnarray}
where both sides are contracted with states in $\Hr$. In the second step we used the BPZ even property of $\mathcal{G}$ and $\mathcal{G}^{-1}$, in the third step we used $\mathcal{G}\mathcal{G}^{-1}=\mathbb{I}$ when operating on $\Hr$. Finally we used the fact that $s_1$ is BPZ odd. Therefore we can compute the variation of the action:
\begin{eqnarray}
\delta_{\mathrm{susy}} S_\mathrm{free} \lineup = \frac{1}{2}\wr(S_1\Psi,Q\Psi) + \frac{1}{2}\wr(\Psi,QS_1\Psi)\nonumber\\
\lineup = \frac{1}{2}\wr(S_1\Psi,Q\Psi) + \frac{1}{2}\wr(\Psi,S_1Q\Psi)\nonumber\\
\lineup = 0.
\end{eqnarray}
The free action is supersymmetric. 

\subsection{Supersymmetry in the Nonlinear Theory}
\label{subsec:susynon}

In the full string field theory, supersymmetry is realized as a nonlinear transformation of the string field: 
\begin{equation}
\delta_{\mathrm{susy}}\Psi = S_1\Psi + S_2(\Psi,\Psi) + S_3(\Psi,\Psi,\Psi)+ \mathrm{higher\ orders}.\label{eq:susysmall}
\end{equation}
The degree even products $S_{n+1}$ will be constructed so that this transformation leaves the action invariant. We will call $S_{n+1}$ {\it supersymmetry products}. Invariance of the action requires the following:
\begin{description}
\item{(S.a)} The supersymmetry products define a symmetry of the $A_\infty$ algebra $\M$ of open superstring field theory. Specifically, if $\S_{n+1}$ is the coderivation corresponding to $S_{n+1}$, the sum
\begin{equation}\S = \sum_{n=0}^\infty \S_{n+1}\end{equation}
commutes with $\M$ 
\begin{equation}\[\S,\M\] = 0.\end{equation}
\item{(S.b)} The supersymmetry products carry the appropriate ghost and picture number so that \eq{susysmall} preserves the ghost and picture number of $\Psi$. 
\item{(S.c)} The supersymmetry products multiply consistently in the small Hilbert space. In particular, we require
\begin{equation}\[\n,\S\] = 0.\end{equation}
\item{(S.d)} The supersymmetry products preserve the Ramond constraint $\mathscr{X}\mathscr{Y} = 1$ when acting on states in $\Hr$. 
\item{(S.e)} The supersymmetry products must be cyclic with respect to the restricted symplectic form:
\begin{equation}\langle \wr|\pi_2\S = 0, \ \ \ \ \mathrm{on}\ T\Hr.\end{equation}
\end{description}
These conditions are closely analogous to those defining the dynamical products $M_{n+1}$. Conditions (S.b), (S.c) and (S.d) imply that the restricted space is closed under multiplication with the supersymmetry products. Note that conditions (S.a), (S.b) and (S.c) are sufficient to imply supersymmetry at the level of the equations of motion, as described in \cite{Ramond}. We additionally require conditions (S.d) and (S.e) to have a supersymmetric action.

Now let us prove that conditions (S.a)-(S.e) imply a symmetry of the action. For this purpose it is helpful to write the action in a form which is closely related to the WZW-like formulation of open superstring field theory \cite{MatsunagaWZW,WBlarge}. We introduce a family of string fields $\Psi(t)\in \Hr$, where $t\in[0,1]$ is an auxiliary parameter, and impose boundary conditions
\begin{equation}\Psi(0) = 0,\ \ \ \Psi(1)=\Psi,\end{equation}
where at $t=1$ we recover the dynamical string field $\Psi$. The action can be written 
\begin{equation}
S = \int_0^1 dt\,\wr\left(\dot{\Psi}(t),\pi_1\M\frac{1}{1-\Psi(t)}\right),\label{eq:smallactiont}
\end{equation}
where $\dot{\Psi}(t) = d\Psi(t)/dt$. The integration over $t$ in \eq{smallactiont} is actually the integral of a total derivative, and if we perform the integral we recover the action as expressed in \eq{smallaction}. In particular, the action only depends on $\Psi(t)$ at $t=1$. The supersymmetry transformation \eq{susysmall} can be expressed as
\begin{equation}
\delta_{\mathrm{susy}}\Psi(t) = \pi_1\S\frac{1}{1-\Psi(t)}.
\end{equation}
Compute
\begin{eqnarray}
\delta_{\mathrm{susy}} S \lineup =  \int_0^1 dt\,\wr\left(\delta_{\mathrm{susy}}\dot{\Psi}(t),\pi_1\M\frac{1}{1-\Psi(t)}\right)+\int_0^1 dt\,\wr\left(\dot{\Psi}(t),\pi_1\M\frac{1}{1-\Psi(t)}\otimes \delta_{\mathrm{susy}}\Psi(t)\otimes \frac{1}{1-\Psi(t)}\right)\nonumber\\
\lineup = 
\int_0^1 dt\,\wr\left(\pi_1\S \frac{1}{1-\Psi(t)}\otimes \dot{\Psi}(t)\otimes \frac{1}{1-\Psi(t)},\pi_1\M\frac{1}{1-\Psi(t)}\right)\nonumber\\
\lineup\ \ \ \ \ \ +\int_0^1 dt\,\wr\left(\dot{\Psi}(t),\pi_1\M\frac{1}{1-\Psi(t)}\otimes \left(\pi_1\S \frac{1}{1-\Psi(t)}\right)\otimes \frac{1}{1-\Psi(t)}\right).\label{eq:susyinv1}
\end{eqnarray}
Note that in computing this we are already assuming conditions (S.b), (S.c) and (S.d), since the supersymmetry variation must be well defined in $\Hr$. Recall that a coderivation $\D$ acts on a group-like element $\frac{1}{1-A}$ as
\begin{equation}\D\frac{1}{1-A} = \frac{1}{1-A}\otimes\left(\pi_1\D\frac{1}{1-A}\right)\otimes\frac{1}{1-A}.\end{equation}
Then the second term in \eq{susyinv1} can be simplified to
\begin{equation}
\int_0^1 dt\,\wr\left(\dot{\Psi}(t),\pi_1\M\S \frac{1}{1-\Psi(t)}\right).\label{eq:susyinv2}
\end{equation}
Moreover, if a coderivation $\D$ is cyclic with respect to $\omega$, we have the relation
\begin{eqnarray}\lineup \omega\left(\pi_1\D\frac{1}{1-A}\otimes B_1\otimes \frac{1}{1-A}\otimes... \otimes \frac{1}{1-A}\otimes B_{n+1}\otimes \frac{1}{1-A},B_{n+2}\right) \nonumber\\
\lineup = -(-1)^{\deg(\D)\deg(B_1)}\omega\left(B_1,\pi_1\D\frac{1}{1-A}\otimes B_2\otimes \frac{1}{1-A}\otimes... \otimes \frac{1}{1-A}\otimes B_{n+2}\otimes \frac{1}{1-A}\right).\label{eq:grpcyc}\end{eqnarray}
Noting $\langle\omega|\pi_2\D = 0$, this can be derived from 
\begin{equation}
\langle\omega|\pi_2 \D \frac{1}{1-A}\otimes B_1\otimes \frac{1}{1-A}\otimes... \otimes \frac{1}{1-A}\otimes B_{n+2}\otimes \frac{1}{1-A} = 0,
\end{equation}
upon expressing the projector $\pi_2$ in the form
\begin{equation}\pi_2 = \inverttriangle (\pi_1\otimes'\pi_1),\triangle\end{equation}
where $\inverttriangle$ is the product and $\triangle$ is the coproduct on the tensor algebra \cite{WB}. Since condition (S.e) implies that $\S$ is cyclic with respect to $\wr$, we can simplify the first term in \eq{susyinv1}:
\begin{eqnarray}
\lineup \int_0^1 dt\,\wr\left(\pi_1\S \frac{1}{1-\Psi(t)}\otimes \dot{\Psi}(t)\otimes \frac{1}{1-\Psi(t)},\pi_1\M\frac{1}{1-\Psi(t)}\right)\nonumber\\
\lineup\ \ \ \ \ \ \ \ \ \ \ \ \ \ =
-\int_0^1 dt\,\wr\left(\dot{\Psi}(t),\pi_1\S\frac{1}{1-\Psi(t)}\otimes\left(\pi_1\M\frac{1}{1-\Psi(t)}\right)\otimes \frac{1}{1-\Psi(t)}\right)\nonumber\\
\lineup\ \ \ \ \ \ \ \ \ \ \ \ \ \ = -\int_0^1 dt\,\wr\left(\dot{\Psi}(t),\pi_1\S\M \frac{1}{1-\Psi(t)}\right).\label{eq:susyinv3}
\end{eqnarray}
Taking \eq{susyinv2} and \eq{susyinv3} together, the variation of the action simplifies to
\begin{eqnarray}
\delta_{\mathrm{susy}} S \lineup = \int_0^1 dt\,\wr\left(\dot{\Psi}(t),\pi_1(-\S\M+\M\S)\frac{1}{1-\Psi(t)}\right)\nonumber\\
\lineup = -\int_0^1 dt\,\wr\left(\dot{\Psi}(t),\pi_1[\S,\M]\frac{1}{1-\Psi(t)}\right),
\end{eqnarray}
which vanishes by condition (S.a). Therefore conditions (S.a)-(S.e) are sufficient to imply a symmetry of the action.

\subsection{Supersymmetry Products}

Now we describe the construction of the supersymmetry transformation. For the time being we will only be interested in implementing conditions (S.a), (S.b) and (S.c), which effectively means that we are constructing a supersymmetry transformation at the level of the equations of motion. In particular, we will not require that the Ramond field satisfies the constraint $\mathscr{X}\mathscr{Y}A=A$, and we will not assume that the dynamical products  $M_{n+1}$ satisfy (M.d) and (M.e). Later we will account for these conditions and specify the supersymmetry transformation satisfying all conditions (S.a)-(S.e). 

The supersymmetry product $S_1$ has components at Ramond number $-1$ and $1$:
\begin{equation}
S_1 = S_1|_{-1} + s_1|_1.
\end{equation}
Both components are degree even and carry ghost number $0$, while $S_1|_{-1}$ carries picture $+1/2$ and $s_1|_1$ carries picture $-1/2$. As in subsection \ref{subsec:freesusy},  we assume that $s_1|_1$ is the Ramond number 1 component of $s_1$. Earlier we chose $S_1|_{-1}=X s_1|_{-1}$, but here we would like to give a more general definition. We postulate that $S_1|_{-1}$ can be expressed in the form
\begin{equation}\S_1|_{-1} = \[\Q,\ssigma_1|_{-1}\],\label{eq:S1m1}\end{equation}
where $\sigma_1|_{-1}$ is a degree odd operator of ghost number $-1$ and picture $+1/2$. The operator $\sigma_1|_{-1}$ is the first example of what we will call a {\it gauge supersymmetry product}. We further assume that $\sigma_1|_{-1}$ satisfies 
\begin{equation}
\[\n,\ssigma_1|_{-1}\] = \s_1|_{-1},\label{eq:s1m1}
\end{equation}
where $s_1|_{-1}$ is the Ramond number $-1$ component of $s_1$. The operator $s_1|_{-1}$ is the first example of what we will call a {\it bare supersymmetry product}. It is clear that $S_1|_{-1}$ will carry one more unit of picture than $s_1|_{-1}$, which is to say that $S_1|_{-1}$ carries picture $+1/2$. We also have the identities 
\begin{equation}\[\Q,\S_1|_{-1}\] = 0,\ \ \ \[\n,\S_1|_{-1}\] = 0.\end{equation}
The first follows from \eq{S1m1} by construction, and the second follows from \eq{S1m1} and \eq{s1m1} after noting that $s_1$ is BRST invariant. Therefore, we have a definition $S_1$ satisfying conditions (S.a), (S.b) and (S.c), which for the moment is our primary concern. Satisfying conditions (S.d) and (S.e) requires a particular choice of contracting homotopy for $\n$ when defining the gauge supersymmetry product from \eq{s1m1}. To reproduce the formula of subsection \ref{subsec:freesusy}, we must choose
\begin{equation}
\sigma_1|_{-1} = \Xi s_1|_{-1}.\label{eq:linsigchoice}
\end{equation}
In this case $S_1$ is consistent with all conditions (S.a)-(S.e).

Next consider the supersymmetry product $S_2$. Condition (S.a) implies it should satisfy
\begin{equation}
\[\Q,\S_2\] + \[\M_2,\S_1\]=0.\label{eq:St2a}
\end{equation}
It is consistent to assume $S_2$ has nonvanishing components at Ramond number $-1$ and $1$: 
\begin{equation}S_{2}=S_2|_{-1} + S_2|_1.
\end{equation}
Both components are degree even and carry ghost number $-1$, while $S_2|_{-1}$ carries picture $+3/2$ and $S_2|_1$ carries picture $+1/2$. We can split \eq{St2a} into components of definite Ramond number: 
\begin{eqnarray}
\[\Q,\S_2|_{-1}\] + \[\M_2|_0,\S_1|_{-1}\] \lineup = 0,\label{eq:St2m1a}\\
\ \[\Q,\S_2|_1\] + \[\m_2|_2,\S_1|_{-1}\] + \[\M_2|_0,\s_1|_1\] \lineup = 0.\label{eq:St21a}
\end{eqnarray}
The strategy is to solve for $S_2|_{-1}$ and $S_2|_1$ by pulling a factor of $\[\Q,\cdot\]$ out of these equations.  Noting 
\begin{equation}\M_2|_0 = \[\Q,\mmu_2|_0\],\ \ \ \S_1|_{-1} = \[\Q,\ssigma_1|_{-1}\],\end{equation}
we can rewrite \eq{St2m1a} and \eq{St21a} as
\begin{eqnarray}
\bigg[\Q,\Big(\S_2|_{-1} - \[\M_2|_0,\ssigma_1|_{-1}\]\Big)\bigg] \lineup = 0,\\
\ \bigg[\Q,\Big(\S_2|_1 - \[\m_2|_2,\ssigma_1|_{-1}\] - \[\s_1|_1,\mmu_2|_0\]\Big)\bigg] \lineup = 0.
\end{eqnarray}
The objects in the commutator with $\Q$ must vanish up $\Q$-exact terms. These terms should be chosen to ensure that $S_2$ is well defined in the small Hilbert space. In this way we find
\begin{eqnarray}
\S_2|_{-1} \lineup =\[\Q,\ssigma_2|_{-1}\] + \[\M_2|_0,\ssigma_1|_{-1}\],\label{eq:St2m1}\\
\S_2|_1 \lineup =  \[\m_2|_2,\ssigma_1|_{-1}\] + \[\s_1|_1,\mmu_2|_0\].\label{eq:St21}
\end{eqnarray}
In the first equation we added a $\Q$-exact term defined by a new gauge supersymmetry product, which we write $\sigma_2|_{-1}$.  In the second equation a $\Q$-exact term is not necessary, since $S_2|_1$ is already in the small Hilbert space:
\begin{eqnarray}
\[\n,\S_2|_1\] = -\[\m_2|_2,\s_1|_{-1}\] + \[\s_1|_1,\m_2|_0\] = \[\s_1,\m_2\]|_1  = 0.
\end{eqnarray}
Let us introduce a bare supersymmetry product $s_2|_{-1}$ satisfying
\begin{equation}\[\n,\ssigma_2|_{-1}\] = \s_2|_{-1}.\label{eq:sig2m1}
\end{equation}
Requiring 
\begin{equation}\[\n,\S_2|_{-1}\] = 0\end{equation}
implies
\begin{eqnarray}
0\lineup =-\[\Q,\s_2|_{-1}\]-\[\M_2|_0,s_1|_{-1}\]\nonumber\\
\lineup = -\bigg[\Q,\Big(\s_2|_{-1}-\[\s_1|_{-1},\mmu_2|_0\]\Big)\bigg].
\end{eqnarray}
Therefore the bare supersymmetry product $s_2|_{-1}$ can be defined
\begin{equation}
\s_2|_{-1} = \[\s_1|_{-1},\mmu_2|_0\],\label{eq:s2m1}
\end{equation}
up to a $\Q$-exact term. However, such a term is not necessary since this definition already implies that $s_2|_{-1}$ is in the small Hilbert space: 
\begin{equation}
\[\n,\s_2|_{-1}\] =  \[\s_1|_{-1},\m_2|_0\] =  \[\s_1,\m_2\]|_{-1} = 0.
\end{equation}
Now that we have the bare supersymmetry product $s_2|_{-1}$, we may define the gauge supersymmetry product $\sigma_2|_{-1}$ with a choice of contracting homotopy for $\n$. This then determines $S_2$ consistent with conditions (S.a), (S.b) and (S.c).

Let us describe the construction to all orders. We introduce an infinite sequence of bare supersymmetry products and gauge supersymmetry products:
\begin{eqnarray}
\mathrm{bare\ supersymmetry\ products}\ \ s_{n+1}|_{-1}:\lineup \ \ \ \ \ \mathrm{degree\ even},
\nonumber\\
\mathrm{gauge\ supersymmetry\ products}\ \ \sigma_{n+1}|_{-1}:\lineup\ \ \ \ \ \mathrm{degree\ odd}.
\end{eqnarray}
We have already described these products when $n=0$ and $n=1$. At higher order, they can be described by generating functions
\begin{eqnarray}
\s|_{-1}(t) \lineup = \sum_{n=0}^\infty t^n\s_{n+1}|_{-1},\\
\ssigma|_{-1}(t) \lineup = \sum_{n=0}^\infty t^n\ssigma_{n+1}|_{-1},
\end{eqnarray}
satisfying the equations
\begin{eqnarray}
\frac{d}{dt}\s|_{-1}(t)\lineup = \[\s|_{-1}(t),\mmu|_0(t)\],\\
\ \[\n,\ssigma|_{-1}(t)\]\lineup = \s|_{-1}(t).\label{eq:sigs}
\end{eqnarray}
Expanding in powers of $t$ gives a recursive definition of these products:
\begin{eqnarray}
\s_{n+2}|_{-1}\lineup = \frac{1}{n+1}\sum_{k=0}^{n}\[\s_{k+1}|_{-1},\mmu_{n-k+2}|_0\],\Vspace\\
\ \[\n,\ssigma_{n+1}|_{-1}\] \lineup = \s_{n+1}|_{-1}.\Vspace\label{eq:nsigs}
\end{eqnarray}
The last equation can be solved with a choice of contracting homotopy for $\n$. The choice does not matter for implementing conditions (S.a), (S.b) and (S.c), but it does matter for conditions (S.d) and (S.e). We will return to this later. With these ingredients we can write the coderivation representing the supersymmetry transformation: 
\begin{equation}\S = [\M,\ssigma|_{-1}] + \s|_1.\Vspace \label{eq:cosusysmallR}\end{equation}
Here we introduce a coderivation $\s|_1$, which represents the $t=1$ value of the generating function 
\begin{equation}
\s|_1(t) = \sum_{n=0}^\infty t^n \s_{n+1}|_1,
\end{equation}
for a sequence of degree even products $s_{n+1}|_1$. The product $s_1|_1$ is the Ramond number $1$ component of $s_1$. The generating function $\s|_1(t)$ is postulated to satisfy
\begin{equation}
\frac{d}{dt} \s|_1(t) = \[\s|_1(t),\mmu|_0(t)\].
\end{equation}
Expanding in powers of $t$ gives a recursive formula for $s_{n+1}|_1$: 
\begin{eqnarray}
\s_{n+2}|_{1} = \frac{1}{n+1}\sum_{k=0}^{n}\[\s_{k+1}|_{1},\mmu_{n-k+2}|_0\].\Vspace
\end{eqnarray}
This gives a construction of the supersymmetry transformation satisfying conditions (S.a), (S.b), and (S.c).

It is not difficult to verify that $\S$ carries the right ghost and picture numbers, so condition (S.b) is satisfied. To verify (S.a) and (S.c), it is useful to express $\s|_1$ and $\s|_{-1}$ in the form 
\begin{eqnarray}
\s|_1 \lineup = \G^{-1} \s_1|_1\G,\label{eq:Gs1}\\
\s|_{-1} \lineup = \G^{-1}\s_1|_{-1}\G,\label{eq:Gsm1}
\end{eqnarray}
using $\G$ from \eq{G}. To check (S.a) we compute 
\begin{equation}
\[\M,\S\] = \[\M,\s|_1\],
\end{equation}
where the first term in $\S$ drops out by $\[\M,\M\] = 0$. Plugging in \eq{GMt} and \eq{Gs1}, we may reexpress this
\begin{equation}
\[\M,\S\] = \G^{-1}\[\Q+\m_2|_2,\s_1|_1\]\G.
\end{equation}
$\s_1|_1$ commutes with $\Q$ and $\[\m_2|_2,\s_1|_1\]$ vanishes by Ramond number counting. Therefore condition (S.a) is satisfied. To check (S.c) we compute 
\begin{equation}
\[\n,\S\] =  -\[\M,\s|_{-1}\] + \[\n,\s|_1\],
\end{equation}
where we used $\[\n,\M\] = 0$ and $\[\n,\ssigma|_{-1}\] = \s|_{-1}$. Plugging in \eq{Gs1},\eq{Gsm1} and \eq{Gn} this becomes
\begin{equation}
\[\n,\S\] = \G^{-1}\Big(-\[\Q+\m_2|_2,\s_1|_{-1}\]+\[\n-\m_2|_0,\s_1|_1\] \Big)\G.
\end{equation}
The commutators with $\Q$ and $\n$ drop out since $s_1$ is BRST invariant and in the small Hilbert space. This leaves
\begin{equation}
\[\n,\S\] = -\G^{-1}\[\m_2,\s_1\]|_1\G,
\end{equation}
which vanishes since $s_1$ is a derivation of the star product. This proves condition (S.c).

Let us explain the relation between the supersymmetry transformation constructed here and the one given in~\cite{Ramond}. The supersymmetry transformation of \cite{Ramond} is characterized by a specific choice of gauge supersymmetry products:
\begin{equation}
\ssigma|_{-1} = \G^{-1}\ssigma_1|_{-1}\G,\label{eq:Rsusy}
\end{equation}
where $\ssigma_1|_{-1}$ is assumed to be the Ramond number $-1$ component of the operator
\begin{equation}
\sigma_1 \equiv \sqrt{2}\oint_{|z|=1} \frac{dz}{2\pi i} \xi\Theta_{a} e^{-\phi/2}(z) \eps_{a}.\label{eq:Rs1}
\end{equation}
This is a special case of the supersymmetry transformation we have been describing. To see this, we must verify
\begin{equation}[\n,\ssigma|_{-1}]=\s|_{-1},\end{equation}
so that \eq{Rsusy} and \eq{Rs1} represents a choice of contracting homotopy for $\n$ in the solution of \eq{sigs}. Compute
\begin{eqnarray}
\[\n,\ssigma|_{-1}\] = \G^{-1}\[\n-\m_2|_0,\ssigma_1|_{-1}\]\G.
\end{eqnarray}
The operator $\sigma_1$ in \eq{Rs1} is a derivation of the star product, and moreover $[\eta,\sigma_1] = s_1$. Thus we can simplify:
\begin{equation}
\[\n,\ssigma|_{-1}\] = \G^{-1}\s_1|_{-1}\G = \s|_{-1},
\end{equation}
which agrees with \eq{nsigs}. An attractive feature of this supersymmetry transformation is that it corresponds a polynomial transformation of the field $\varphi$ in the Chern-Simons-like equations \eq{CSEOM}. Moreover, the transformation of $\varphi$ requires no picture changing insertions which break conformal invariance, which is convenient for the analysis of analytic solutions. However, the supersymmetry transformation of \cite{Ramond} is not a symmetry of the action since it does not implement conditions (S.d) and (S.e).

\subsection{Cyclic Ramond Number Decomposition}
\label{subsec:susycyc}

To realize supersymmetry in the action we must make a specific choice of gauge supersymmetry products. Assuming that the dynamical products $M_{n+1}$ are given as in section \ref{sec:Ainf}, we claim the proper choice is
\begin{equation}
\sigma_{m+1}|_{-1} = \Xi s_{m+1}|_{-1}.\label{eq:gss}
\end{equation}
In this subsection our task is to use this form of $\sigma_{m+1}|_{-1}$ to express the supersymmetry products in components of definite cyclic Ramond number. In this process we will see that condition (S.d) is satisfied. The proof of cyclicity of the supersymmetry products in the next subsection will then proceed by demonstrating cyclicity of the cyclic Ramond number components. 

Assuming the gauge products $\mu_{n+2}|_0$ are defined as in section \ref{sec:Ainf}, the cohomomorphism $\G^{-1}$ takes a special form when it produces one Ramond output \cite{RWaction}:
\begin{equation}
\pi_1^1\G^{-1}  =\pi_1^1\Big(\mathbb{I}_{T\H} - \Xi\m_2|_0\Big).\label{eq:RGinv}
\end{equation}
From this it follows that the gauge supersymmetry products also take a special form with one Ramond output: 
\begin{eqnarray}\pi_1^1\ssigma|_{-1}\lineup  = \pi_1^1 \Xi \s|_{-1} \nonumber\\
\lineup = \Xi\, \pi_1^1 \G^{-1}\s_1|_{-1}\G\nonumber\\
\lineup = \pi_1^1\Xi\s_1|_{-1} \G.\label{eq:Rsig}
\end{eqnarray}
In the final step we used $\Xi^2 = 0$. Similar computations show that \cite{RWaction}
\begin{eqnarray}
\pi_1^1 \m|_0 \lineup = \pi_1^1\m_2|_0\G,\label{eq:Rm0}\\
\pi_1^1 \M|_0 \lineup = \pi_1^1\Big(\Q +\, X \m|_0\Big),\label{eq:RM0}\\
\pi_1^1 \m|_2 \lineup = 0.\label{eq:Rm2}
\end{eqnarray} 
The last two relations reexpress \eq{specprod}. 

We now use these formulas to compute the cyclic Ramond number decomposition of $\S$. Since $S_{n+1}$ has components at Ramond number $-1$ and $1$, potentially it can have components at cyclic Ramond number $1$ and~$3$:
\begin{equation}S_{n+1} = S_{n+1}|^1 + S_{n+1}|^3.\end{equation}
First consider the cyclic Ramond number $3$ component. We know that $S_{n+1}|_3^3$ vanishes since $S_{n+1}$ does not carry Ramond number $3$. Cyclicity should then imply that $S_{n+1}|_1^3$ also vanishes. However, this fact is nontrivial. We can compute $S_{n+1}|_1^3$ using \eq{slashrrp2}:
\begin{equation}S_{n+1}|_1^3\pi_{n+1} = \pi_1^1\S \pi_{n+1}^2.\end{equation}
Plugging in \eq{cosusysmallR} gives 
\begin{eqnarray}
S_{n+1}|_1^3\pi_{n+1} \lineup = \pi_1^1\Big([\M,\ssigma|_{-1}]+ \s|_1\Big)\pi_{n+1}^2\nonumber\\
\lineup = \pi_1^1\Big([\m|_2,\ssigma|_{-1}]+ \s|_1\Big)\pi_{n+1}^2\nonumber\\
\lineup = \pi_1^1\Big(\ssigma|_{-1}\m|_2+\s|_1\Big)\pi_{n+2}^2,
\end{eqnarray}
where we used \eq{Rm2} to drop one term from the commutator. Next we plug in \eq{Rsig} for $\ssigma|_{-1}$ and expand $\s|_1$ using \eq{RGinv} to find
\begin{eqnarray}
S_{n+1}|_1^3\pi_{n+1} \lineup = \pi_1^1\Big( \Xi\s_1|_{-1}\G \m|_2 + \s_1|_1\G -\Xi\m_2|_0\s_1|_1\G\Big)\pi_{n+1}^2.
\end{eqnarray}
The first term can be further simplified using $\m|_2 = \G^{-1}\m_2|_2\G$. The second term drops out since $s_1|_1$ produces only an NS output. Therefore
\begin{eqnarray}
S_{n+1}|_1^3\pi_{n+1} \lineup = \Xi \,\pi_1^1\Big(\s_1|_{-1}\m_2|_2 - \m_2|_0\s_1|_1\Big)\G\,\pi_{n+1}^2\nonumber\\
\lineup = \Xi \, \pi_1^1\Big( \[\s_1|_{-1},\m_2|_2\] + \[\s_1|_1,\m_2|_0\]\Big)\G\,\pi_{n+1}^2\nonumber\\
\lineup = \Xi \, \pi_1^1[\s_1,\m_2]\big|_1 \G\, \pi_{n+1}^2\nonumber\\
\lineup = 0.
\end{eqnarray}
In the second step substituted commutators since $m_2|_2$ and $s_1|_1$ only produce NS outputs. Finally we used that $s_1$ is a derivation of the star product. Therefore the cyclic Ramond number $3$ component vanishes. 

By process of elimination, this means that the supersymmetry products only carry cyclic Ramond number $1$. In particular, there can only be one Ramond state in the input and output of the supersymmetry transformation. When the supersymmetry products produce an NS state, they take the form 
\begin{eqnarray}
S_{n+1}|_{1}^1\pi_{n+1}   \lineup =  \pi_1^0\S\pi_{n+1}^1\nonumber\\
\lineup = \pi_1^0\Big([\M,\ssigma|_{-1}] +\s|_1\Big)\pi_{n+1}^1\nonumber\\
\lineup = \pi_1^0\Big([\m|_2,\ssigma|_{-1}] +\s|_1\Big)\pi_{n+1}^1,\label{eq:St11}
\end{eqnarray}
and when they produce a Ramond state, 
\begin{eqnarray}
S_{n+1}|_{-1}^1\pi_{n+1}   \lineup =  \pi_1^1\S\pi_{n+1}^0\nonumber\\
\lineup = \pi_1^1\Big([\M,\ssigma|_{-1}] +\s|_1\Big)\pi_{n+1}^0\nonumber\\
\lineup = \pi_1^1\[\M|_0,\ssigma|_{-1}\]\pi_{n+1}^0.
\end{eqnarray}
Using \eq{RM0} and \eq{Rsig} this can be further expressed
\begin{eqnarray}
S_{n+1}|_{-1}^1\pi_{n+1}\lineup = \pi_1^1\Big((\Q+ X\m|_0)\ssigma|_{-1} +\ssigma|_{-1}\M|_0 \Big)\pi_{n+1}^0\nonumber\\
\lineup = \pi_1^1 \Big(\Q\Xi\s_1|_{-1}\G+ X\m|_0 \ssigma|_{-1}+ \Xi\s_1|_{-1}\Q\G\Big)\pi_{n+1}^0\nonumber\\
\lineup =X\, \pi_1^1\Big(\s_1|_{-1} \G +\m|_0 \ssigma|_{-1}\Big)\pi_{n+1}^0  - \Xi\,\pi_1^1 [\Q,\s_1|_{-1}] \G\pi_{n+1}.
\end{eqnarray}
The last term drops out since $Q$ and $s_1$ commute. Now use \eq{RGinv} to insert a factor of $\G^{-1}$ in front of $\s_1|_{-1}$ in the first term:
\begin{eqnarray}
S_{n+1}|_{-1}^1\pi_{n+1}\lineup = X\, \pi_1^1\Big(\G^{-1}\s_1|_{-1} \G +\Xi\m_2|_0 \s_1|_{-1}\G +\m|_0 \ssigma|_{-1}\Big)\pi_{n+1}^0\nonumber\\
\lineup = X\, \pi_1^1\Big(\s|_{-1} +\Xi\m_2|_0 \s_1|_{-1}\G +\m|_0 \ssigma|_{-1}\Big)\pi_{n+1}^0.
\end{eqnarray}
In the second term we can switch the order of $\s_1|_{-1}$ and $\m_2|_0$ since $s_1$ is a derivation of the star product. Inserting a factor of $\G\G^{-1}$ then gives
\begin{eqnarray}
S_{n+1}|_{-1}^1\pi_{n+1} \lineup = X\, \pi_1^1\Big(\s|_{-1} +\Xi\s_1|_{-1} \G\G^{-1}\m_2|_0 \G +\m|_0 \ssigma|_{-1}\Big)\pi_{n+1}^0\nonumber\\
\lineup = X\, \pi_1^1\Big(\s|_{-1} +\ssigma|_{-1} \m|_0 +\m|_0 \ssigma|_{-1}\Big)\pi_{n+1}^0\nonumber\\
\lineup =  X\, \pi_1^1\Big(\s|_{-1} +\[\m|_0,\ssigma|_{-1}\]\Big)\pi_{n+1}^0.
\end{eqnarray}
Taking the NS and R outputs together, we therefore have 
\begin{eqnarray}
S_{n+1}\pi_{n+1}\lineup = (S_{n+1}|^1_{-1} + S_{n+1}|^1_1)\pi_1\nonumber\\
\lineup = X\, \pi_1^1\Big(\[\ssigma|_{-1}, \m|_0\]+\s|_{-1} \Big)\pi_{n+1}^0+\pi_1^0\Big([\m|_2,\ssigma|_{-1}] +\s|_1\Big)\pi_{n+1}^1.
\end{eqnarray}
It is natural to combine $\s|_{-1}$ and $\s|_1$ into a single object: 
\begin{eqnarray}
\s\lineup  \equiv \s|_{-1}+\s|_1 = \G^{-1}\s_1\G .\label{eq:sst}
\end{eqnarray}
The supersymmetry products can then be expressed
\begin{equation}
\S = \mathcal{G} \Big([\m,\ssigma|_{-1}] + \s\Big)\Big|^1.\label{eq:ScycD}
\end{equation}
Note that the Ramond output is proportional to $X$, and is therefore consistent with the condition $\mathscr{X}\mathscr{Y} = 1$ on the Ramond string field. Therefore the supersymmetry products realize condition (S.d). 

\subsection{Proof of Cyclicity}

We are now ready to prove cyclicity: 
\begin{equation}
\langle \wr|\pi_2\S  = 0\ \ \mathrm{on}\ \ \Hr.\label{eq:ScycHtres}
\end{equation}
From \eq{ScycD} we see that the factor of $\mathcal{G}^{-1}$ in the restricted symplectic form cancels a factor of $\mathcal{G}$ in the supersymmetry products. Therefore cyclicity is equivalent to
\begin{equation}
\langle \omega_S|\pi_2 \Big([\m,\ssigma|_{-1}] +\s\Big)\Big|^1=0\ \ \mathrm{on}\ \ \Hr.
\end{equation}
The argument goes more easily if at intermediate steps we allow ourselves to contract with states in the large Hilbert space. Therefore we will prove the stronger relation 
\begin{equation}
\langle \omega_L|\pi_2 \Big([\m,\ssigma|_{-1}] +\s\Big)\Big|^1=0,
\end{equation}
which holds contracted with arbitrary states in $\H$. The restriction to cyclic Ramond number 1 can be implemented by operating with the projector $\pi^1$:
\begin{equation}
\langle \omega_L|\pi_2\Big([\m,\ssigma|_{-1}] +\s\Big)\pi^1=0.
\end{equation}
Next we separate $\s$ and $\m$ into components of definite Ramond number, and commute the projector $\pi^1$ through to operate on $\pi_2$. This produces
\begin{equation}
\langle\omega_L|\pi_2^0\Big(\s|_1+\[\m|_2,\ssigma|_{-1}\]\Big) + \langle\omega_L|\pi_2^2\Big(\s|_{-1} +\[\m|_0,\ssigma|_{-1}\] \Big)  =0.
\label{eq:coScyc}
\end{equation}
Our task is to show that the two terms cancel. 

Let us start with the first term. Since $\ssigma|_{-1}$ necessarily produces a Ramond output, one piece of the commutator vanishes against $\pi_2^0$, which only accepts NS inputs. Also, it follows from the construction of the NS open superstring field theory \cite{WittenSS} that the cohomomorphism $\G^{-1}$ is cyclic when it produces only NS outputs \cite{RWaction}. In particular this means  
\begin{equation}
\langle \omega_L|\pi_2^0 \G^{-1} = \langle\omega_L|\pi_2^0.
\end{equation}
Therefore the first term in \eq{coScyc} simplifies to 
\begin{equation}
\langle \omega_L|\pi_2^0 \Big(\s_1|_1\G  + \m_2|_2\G\ssigma|_{-1}\Big),
\end{equation}
where we substituted $\s|_1 = \G^{-1}\s_1|_1\G$ and $\m|_2 = \G^{-1}\m_2|_2\G$.

Evaluating the second term in \eq{coScyc} requires more work. We represent the projector $\pi_2^2$ in terms of the product and coproduct \cite{WB,RWaction}
\begin{equation}\pi_2^2 = \inverttriangle(\pi_1^1\otimes'\pi_1^1)\triangle,\end{equation}
and act the coproduct to the right. This gives
\begin{eqnarray}
 \langle\omega_L|\pi_2^2\Big(\s|_{-1} +\[\m|_0,\ssigma|_{-1}\] \Big) \lineup = \langle \omega_L|\inverttriangle\bigg[\big(\pi_1^1\s|_{-1}\big)\otimes'\pi_1^1 + \pi_1^1\otimes'\big(\pi_1^1\s|_{-1}\big)\nonumber\\
 \lineup\ \ \ \ \ \ \ \ \ \ \ \ \ \ \ 
 +\Big(\pi_1^1\[\m|_0,\ssigma|_{-1}\]\Big)\otimes'\pi_1^1 + \pi_1^1\otimes'\Big(\pi_1^1\[\m|_0,\ssigma|_{-1}\]\Big)\bigg]\triangle.
\end{eqnarray}
We can evaluate the action of $\pi_1^1$ on the above coderivations using \eq{RGinv}-\eq{Rm2}: 
\begin{eqnarray}
 \langle\omega_L|\pi_2^2\Big(\s|_{-1} +\[\m|_0,\ssigma|_{-1}\] \Big) \lineup = \langle \omega_L|\inverttriangle\bigg[\Big(\pi_1^1(\s_1|_{-1} -\Xi\m_2|_0\s_1|_{-1})\G\Big)\otimes'\pi_1^1 + \pi_1^1\otimes'\Big(\pi_1^1(\s_1|_{-1} -\Xi\m_2|_0\s_1|_{-1})\G\Big)\nonumber\\
 \lineup\ \ \ \ \ \ \ \ \ \ \ \ \ 
 +\Big(\pi_1^1\m_2|_0\G\ssigma|_{-1}\Big)\otimes'\pi_1^1 + \pi_1^1\otimes'\Big(\pi_1^1\m_2|_0\G\ssigma|_{-1}\Big)\nonumber\\
  \lineup\ \ \ \ \ \ \ \ \ \ \ \ \ 
 +\Big(\pi_1^1\Xi\s_1|_{-1}\m_2|_0 \G \Big)\otimes'\pi_1^1 + \pi_1^1\otimes'\Big(\pi_1^1\Xi\s_1|_{-1}\m_2|_0 \G\Big)\bigg]\triangle.
\end{eqnarray}
Note that the last two terms cancel against the first two terms using 
\begin{equation}
\[\s_1|_{-1},\m_2|_0\] = [\s_1,\m_2]|_{-1}=0,
\end{equation}
so this simplifies to
\begin{eqnarray}
 \langle\omega_L|\pi_2^2\Big(\s|_{-1} +\[\m|_0,\ssigma|_{-1}\] \Big) \lineup = \langle \omega_L|\inverttriangle\bigg[\Big(\pi_1^1\s_1|_{-1}\G\Big)\otimes'\pi_1^1 + \pi_1^1\otimes'\Big(\pi_1^1\s_1|_{-1}\G\Big)\nonumber\\
 \lineup\ \ \ \ \ \ \ \ \ \ \ \ \ 
 +\Big(\pi_1^1\m_2|_0\G\ssigma|_{-1}\Big)\otimes'\pi_1^1 + \pi_1^1\otimes'\Big(\pi_1^1\m_2|_0\G\ssigma|_{-1}\Big)\bigg]\triangle.
\end{eqnarray}
We expand further by writing
\begin{equation}
\pi_1^1 = \pi_1^1\G^{-1}\G = \pi_1^1\G -\pi_1^1\Xi\m_2|_0\G,
\end{equation}
so that
\begin{eqnarray}
 \langle\omega_L|\pi_2^2\Big(\s|_{-1} +\[\m|_0,\ssigma|_{-1}\] \Big) \lineup = \langle \omega_L|\inverttriangle\bigg[\big(\pi_1^1\s_1|_{-1}\G)\otimes'\big(\pi_1^1\G\big) - \big(\pi_1^1\s_1|_{-1}\G)\otimes'\big(\pi_1^1\Xi\m_2|_0\G\big) \nonumber\\
\lineup\ \ \ \ \ \ \ \ \ \ \ \ \ 
+ \big(\pi_1^1\G\big)\otimes'\big(\pi_1^1\s_1|_{-1}\G)- \big(\pi_1^1\Xi\m_2|_0\G\big) \otimes' \big(\pi_1^1\s_1|_{-1}\G\big)\Vspace\nonumber\\
\lineup\ \ \ \ \ \ \ \ \ \ \ \ \ 
 +\big(\pi_1^1\m_2|_0\G\ssigma|_{-1}\big)\otimes'\big(\pi_1^1\G\big)-\big(\pi_1^1\m_2|_0\G\ssigma|_{-1}\big)\otimes'\big(\pi_1^1\Xi\m_2|_0\G\big)\Vspace\nonumber\\
 \lineup\ \ \ \ \ \ \ \ \ \ \ \ \
  + \big(\pi_1^1\G\big)\otimes'\big(\pi_1^1\m_2|_0\G\ssigma|_{-1}\big)-\big(\pi_1^1\Xi\m_2|_0\G\big)\otimes'\big(\pi_1^1\m_2|_0\G\ssigma|_{-1}\big)\bigg]\triangle.\nonumber\\\label{eq:pf1}
\end{eqnarray}
Using the BPZ even property of $\Xi$ we can write the second term
\begin{eqnarray}
\langle\omega_L|\inverttriangle\bigg[-\big(\pi_1^1\s_1|_{-1}\G)\otimes'\big(\pi_1^1\Xi\m_2|_0\G\big)\bigg]\triangle\lineup=
\langle\omega_L|\inverttriangle\bigg[-\big(\pi_1^1\Xi\s_1|_{-1}\G)\otimes'\big(\pi_1^1\m_2|_0\G\big)\bigg]\triangle\nonumber\\
\lineup =\langle\omega_L|\inverttriangle\bigg[-\big(\pi_1^1\ssigma|_{-1})\otimes'\big(\pi_1^1\m_2|_0\G\big)\bigg]\triangle.
\end{eqnarray}
The same manipulation applies to the fourth term in opposite order. Therefore
\begin{eqnarray}
 \langle\omega_L|\pi_2^2\Big(\s|_{-1} +\[\m|_0,\ssigma|_{-1}\] \Big) \lineup = \langle \omega_L|\inverttriangle\bigg[\big(\pi_1^1\s_1|_{-1}\G)\otimes'\big(\pi_1^1\G\big) - \big(\pi_1^1\ssigma|_{-1})\otimes'\big(\pi_1^1\m_2|_0\G\big) \nonumber\\
\lineup\ \ \ \ \ \ \ \ \ \ \ \ \ 
+ \big(\pi_1^1\G\big)\otimes'\big(\pi_1^1\s_1|_{-1}\G)+ \big(\pi_1^1\m_2|_0\G\big) \otimes' \big(\pi_1^1\ssigma|_{-1}\big)\Vspace\nonumber\\
\lineup\ \ \ \ \ \ \ \ \ \ \ \ \ 
 +\big(\pi_1^1\m_2|_0\G\ssigma|_{-1}\big)\otimes'\big(\pi_1^1\G\big)-\big(\pi_1^1\m_2|_0\G\ssigma|_{-1}\big)\otimes'\big(\pi_1^1\Xi\m_2|_0\G\big)\Vspace\nonumber\\
 \lineup\ \ \ \ \ \ \ \ \ \ \ \ \
  + \big(\pi_1^1\G\big)\otimes'\big(\pi_1^1\m_2|_0\G\ssigma|_{-1}\big)-\big(\pi_1^1\Xi\m_2|_0\G\big)\otimes'\big(\pi_1^1\m_2|_0\G\ssigma|_{-1}\big)\bigg]\triangle.\nonumber\\
\end{eqnarray}
Using the BPZ even property of $\Xi$ we can write the sixth term
\begin{eqnarray}
\langle\omega_L|\!\inverttriangle\!\bigg[-\big(\pi_1^1\m_2|_0\G\ssigma|_{-1}\big)\otimes'\big(\pi_1^1\Xi\m_2|_0\G\big)\bigg]\triangle\lineup=
\langle\omega_L|\!\inverttriangle\!\bigg[-\big(\pi_1^1\Xi\m_2|_0\G\ssigma|_{-1}\big)\otimes'\big(\pi_1^1\m_2|_0\G\big)\bigg]\triangle\nonumber\\
\lineup = \langle\omega_L|\!\inverttriangle\!\bigg[\big(\pi_1^1(\G^{-1}-\mathbb{I}_{T\H})\G\ssigma|_{-1}\big)\otimes'\big(\pi_1^1\m_2|_0\G\big)\bigg]\triangle\nonumber\\ 
\lineup = \langle\omega_L|\!\inverttriangle\!\bigg[\big(\pi_1^1\ssigma|_{-1}\big)\otimes'\big(\pi_1^1\m_2|_0\G\big)-\big(\pi_1^1\G\ssigma|_{-1}\big)\otimes'\big(\pi_1^1\m_2|_0\G\big)\bigg]\triangle.\nonumber\\
\end{eqnarray}
The same manipulation applies to the eighth term in opposite order. Thus we find
\begin{eqnarray}
 \langle\omega_L|\pi_2^2\Big(\s|_{-1} +\[\m|_0,\ssigma|_{-1}\] \Big) \lineup = \langle \omega_L|\inverttriangle\bigg[\big(\pi_1^1\s_1|_{-1}\G)\otimes'\big(\pi_1^1\G\big) - \big(\pi_1^1\ssigma|_{-1})\otimes'\big(\pi_1^1\m_2|_0\G\big) \nonumber\\
\lineup\ \ \ \ \ \ \ \ \ \ \ \ \ 
+ \big(\pi_1^1\G\big)\otimes'\big(\pi_1^1\s_1|_{-1}\G)+ \big(\pi_1^1\m_2|_0\G\big) \otimes' \big(\pi_1^1\ssigma|_{-1}\big)\Vspace\nonumber\\
\lineup\ \ \ \ \ \ \ \ \ \ \ \ \ 
 +\big(\pi_1^1\m_2|_0\G\ssigma|_{-1}\big)\otimes'\big(\pi_1^1\G\big)+\big(\pi_1^1\ssigma|_{-1}\big)\otimes'\big(\pi_1^1\m_2|_0\G\big)\Vspace\nonumber\\
\lineup\ \ \ \ \ \ \ \ \ \ \ \ \  
 -\big(\pi_1^1\G\ssigma|_{-1}\big)\otimes'\big(\pi_1^1\m_2|_0\G\big)  + \big(\pi_1^1\G\big)\otimes'\big(\pi_1^1\m_2|_0\G\ssigma|_{-1}\big)\Vspace
 \nonumber\\
\lineup\ \ \ \ \ \ \ \ \ \ \ \ \  
-\big(\pi_1^1\m_2|_0\G\big)\otimes'\big(\pi_1^1\ssigma|_{-1}\big)  + \big(\pi_1^1\m_2|_0\G\big)\otimes'\big(\pi_1^1\G\ssigma|_{-1}\big)\bigg]\triangle.\nonumber\\
\end{eqnarray}
There is some cancellation:
\begin{eqnarray}
 \langle\omega_L|\pi_2^2\Big(\s|_{-1} +\[\m|_0,\ssigma|_{-1}\] \Big) \lineup = \langle \omega_L|\inverttriangle\bigg[\big(\pi_1^1\s_1|_{-1}\G)\otimes'\big(\pi_1^1\G\big) + \big(\pi_1^1\G\big)\otimes'\big(\pi_1^1\s_1|_{-1}\G)\Vspace\nonumber\\
\lineup\ \ \ \ \ \ \ \ \ \ \ \ \ 
 +\big(\pi_1^1\m_2|_0\G\ssigma|_{-1}\big)\otimes'\big(\pi_1^1\G\big)
 -\big(\pi_1^1\G\ssigma|_{-1}\big)\otimes'\big(\pi_1^1\m_2|_0\G\big)\Vspace\nonumber\\
\lineup\ \ \ \ \ \ \ \ \ \ \ \ \  
+ \big(\pi_1^1\G\big)\otimes'\big(\pi_1^1\m_2|_0\G\ssigma|_{-1}\big) + \big(\pi_1^1\m_2|_0\G\big)\otimes'\big(\pi_1^1\G\ssigma|_{-1}\big)\bigg]\triangle.\nonumber\\
\end{eqnarray}
The structure of these terms is such that we can pull the coproduct back to the left towards the product and replace again with $\pi_2^2$. This leaves
\begin{equation}
 \langle\omega_L|\pi_2^2\Big(\s|_{-1} +\[\m|_0,\ssigma|_{-1}\] \Big) = \langle \omega_L|\pi_2^2\Big(\s_1|_{-1}\G +\m_2|_0\G\ssigma|_{-1}\Big).
\end{equation}
Adding the first and second terms in \eq{coScyc} then gives
\begin{eqnarray}
\langle\omega_L|\pi_2\Big(\s+[\m,\ssigma|_{-1}]\Big)\pi^1\lineup = \langle \omega_L|\pi_2^0 \Big(\s_1|_1\G  + \m_2|_2\G\ssigma|_{-1}\Big) +
\langle \omega_L|\pi_2^2\Big(\s_1|_{-1}\G +\m_2|_0\G\ssigma|_{-1}\Big) \nonumber\\
\lineup =
\langle \omega_L|\pi_2 \Big(\s_1|_1\G  + \m_2|_2\G\ssigma|_{-1}\Big)\pi^1 +
\langle \omega_L|\pi_2\Big(\s_1|_{-1}\G +\m_2|_0\G\ssigma|_{-1}\Big)\pi^1\nonumber\\
\lineup =
\langle \omega_L|\pi_2 \Big(\s_1\G  + \m_2\G\ssigma|_{-1}\Big)\pi^1\nonumber\\
\lineup = 0,
\end{eqnarray}
which vanishes since both $s_1$ and $m_2$ are cyclic with respect to the large Hilbert space symplectic form. This completes the proof of cyclicity. Therefore the supersymmetry products  satisfy all conditions (S.a)-(S.e) required to define a symmetry of the action.

\section{Supersymmetry Algebra}
\label{sec:susyalg}

Now that we have the supersymmetry transformation, it is interesting to investigate the form of the supersymmetry algebra. On general grounds we expect the supersymmetry algebra to appear as
\begin{equation}
[\delta_\mathrm{susy}',\delta_\mathrm{susy}] \Psi = -2P_1\Psi + \mathrm{trivial\ terms},\label{eq:susyalg}
\end{equation}
where $P_1$ is the momentum operator 
\begin{equation}
P_1 \equiv \eps_{a} (C\Gamma^\mu)_{ab} \eps'_{b} \oint_{|z|=1}\frac{dz}{2\pi i} i\d X_\mu(z),
\end{equation}
and $\eps_{a}$ and $\eps_{a}'$ are the parameters defining the supersymmetry transformations $\delta_{\mathrm{susy}}$ and $\delta_{\mathrm{susy}}'$, respectively. We will use a prime to denote objects defined with the primed parameter $\eps_{a}'$. The supersymmetry algebra may in addition contain a gauge transformation or symmetry transformation which vanishes on-shell. These transformations act trivially on physical observables.

It is clear that the commutator of supersymmetry transformations will be a nonlinear function of the string field. The momentum operator only acts on one string field, so the remaining nonlinear terms must be a combination of gauge transformations and symmetries which vanish on-shell. For a generic supersymmetric field theory, determining the explicit form of these transformations may be difficult. In our case, with some motivation from the deformation theory of $A_\infty$ algebras, we can anticipate that they will take a fairly specific form. Consider a deformation of an $A_\infty$ algebra $\M\to\M+\delta \M$ with infinitesimal $\delta\M$. The deformation produces a new $A_\infty$ algebra if $\delta\M$ satisfies
\begin{equation}[\M,\delta\M] = 0.\label{eq:HochM}\end{equation}
Since $\M$ is a nilpotent object, this condition defines a cohomology, called the Hochschild cohomology. A trivial element of the Hochschild cohomology takes the form
\begin{equation}\delta\M = [\M,\bm{\Lambda}].\end{equation}
These are precisely the deformations of $\M$ which can be implemented by a field redefinition.\footnote{Following \cite{MoellerSachs}, it is expected that nontrivial elements of the Hochschild cohomology should correspond to closed string states.} Analogously, we may consider deformations of a symmetry ${\bf V}$ of an $A_\infty$ algebra. Since a symmetry satisfies $[\M,{\bf V}] = 0$, a deformation ${\bf V} \to {\bf V} + \delta{\bf V}$ will continue to be a symmetry if 
\begin{equation}[\M,\delta{\bf V}] = 0. \label{eq:HochD}\end{equation}
Note that the ghost number of $\delta {\bf V}$ is decreased by $1$ relative to $\delta\M$. A trivial deformation of the symmetry should be represented by a trivial element of the cohomology:
\begin{equation}\delta{\bf V} = [\M,\T].\end{equation}
With this understanding, it is natural to expect that the commutator of supersymmetry transformations may not exactly produce the momentum operator, but a symmetry which is equivalent to the momentum operator in the cohomology of $[\M,\cdot]$.   Expressed in terms of coderivations, this means that the supersymmetry algebra will take the form
\begin{equation}
\[\S,\S'\] = -2\P_1 + \[\M,\T\],\label{eq:cosusyalg}
\end{equation}
for some degree odd coderivation $\T$. To understand what this implies, compute
\begin{eqnarray}
[\delta_\mathrm{susy}',\delta_\mathrm{susy}]\Psi \lineup = \pi_1\[\S,\S'\]\frac{1}{1-\Psi} \nonumber \\
\lineup = \pi_1(-2\P_1+\[\M,\T\])\frac{1}{1-\Psi}\nonumber\\
\lineup =  -2P_1\Psi +\left(\pi_1\M\frac{1}{1-\Psi}\otimes\left(\pi_1\T\frac{1}{1-\Psi}\right)\otimes \frac{1}{1-\Psi}\right)+\left(\pi_1\T \frac{1}{1-\Psi}\otimes\left(\pi_1\M\frac{1}{1-\Psi}\right)\otimes \frac{1}{1-\Psi}\right).\ \ \ \ \ \ \ 
\end{eqnarray}
The second term above represents an infinitesimal gauge transformation of $\Psi$ with a gauge parameter
\begin{equation}
\pi_1\T \frac{1}{1-\Psi}.\label{eq:susyalg_gauge}
\end{equation}
The third term vanishes assuming the equations of motion
\begin{equation}
 \pi_1\M \frac{1}{1-\Psi} = 0,
\end{equation}
and therefore represents a symmetry which vanishes on-shell.

Therefore our main task is to compute $\T$. For the above interpretation of the supersymmetry algebra to be consistent, we must require the following properties: 
\begin{description}
\item{(T.a)} $\T$ must satisfy \eq{cosusyalg}. 
\item{(T.b)} The products of $\T$ must carry the appropriate ghost and picture number so that \eq{susyalg_gauge} is an allowed gauge transformation of the dynamical string field.
\item{(T.c)} The products of $\T$ multiply consistently in the small Hilbert space. In particular, we require
\begin{equation}\[\n,\T\] = 0.\end{equation}
\item{(T.d)} The products of $\T$ preserve the Ramond constraint $\mathscr{X}\mathscr{Y} = 1$ when acting on states in $\Hr$.
\item{(T.e)} $\T$ must be cyclic with respect to the restricted symplectic form:
\begin{equation}\langle \wr|\pi_2\T = 0, \ \ \ \ \mathrm{on}\ T\Hr.\end{equation}
\end{description}
Conditions (T.b), (T.c) and (T.d) imply that the restricted space is closed under multiplication with the products of $\T$, so that in particular \eq{susyalg_gauge} is a well-defined gauge transformation. Condition (T.e) is required so that $[\M,\T]$ is a cyclic coderivation, and therefore generates a symmetry of the action.

\subsection{Computation of Trivial Term in Supersymmetry Algebra}

As a first step we will give a definition of $\T$ satisfying conditions (T.a), (T.b) and (T.c), leaving conditions (T.d) and (T.e) for the next subsection. We temporarily ignore the constraint on the Ramond string field and the conditions (M.d)-(M.e) and (S.d)-(S.e) for the dynamical and supersymmetry products. 

It is useful to introduce the operators
\begin{eqnarray}
\varpi_1 \lineup \equiv -\frac{1}{\sqrt{2}}\eps_{a} (C\Gamma^\mu)_{ab} \eps'_{b} \oint_{|z|=1}\frac{dz}{2\pi i} \psi_\mu \xi e^{-\phi}(z),\\ 
p_1 \lineup  \equiv \frac{1}{\sqrt{2}}\eps_{a} (C\Gamma^\mu)_{ab} \eps'_{b} \oint_{|z|=1}\frac{dz}{2\pi i} \psi_\mu e^{-\phi}(z).
\end{eqnarray}
$\varpi_1$ is degree odd, ghost number $-1$ and picture $0$, and $p_1$ is degree even, ghost number $0$, and picture $-1$.  We can think of $\varpi_1$ as a ``momentum gauge product" and $p_1$ a ``momentum bare product." We have the relations
\begin{eqnarray}
\ [\Q,\P_1] = 0,\lineup \ \ \ \ \ \, [\Q,\bm{\varpi}_1] = \P_1,\ \ \ \  \,[\Q,\p_1] = 0,\\
\ [\n,\P_1] = 0,\lineup\ \ \ \ \ \ [\n,\bm{\varpi}_1] = \p_1,\ \ \ \ \ \,[\n,\p_1] = 0,
\end{eqnarray}
and
\begin{eqnarray}
\  [\m_2,\P_1] = 0,\lineup\ \ \ \ \, [\m_2,\p_1] = 0,\ \ \ \ \,[\m_2,\bm{\varpi}_1]= 0,\\
\langle\omega_L|\pi_2\P_1=0,\lineup\ \ \ \langle\omega_L|\pi_2\p_1= 0,\ \ \ \langle\omega_L|\pi_2\bm{\varpi}_1 = 0.
\end{eqnarray}
which follow from the fact that $P_1,\varpi_1$ and $p_1$ are zero modes of weight 1 primaries. The operator $p_1$ appears in the ``supersymmetry algebra" generated by $s_1$:
\begin{equation}[s_1,s_1']=-2p_1.\label{eq:m1susyalg}\end{equation}
This is not quite a supersymmetry algebra since $p_1$ is not the standard momentum operator.

To find $\T$ we start by using \eq{cosusysmallR} to compute the commutator 
\begin{equation}
\[\S,\S'\] = \bigg[\,\M,\bigg( \[\s|_1,\ssigma|_{-1}'\] +\[\ssigma|_{-1},\s|_1'\]+\[\ssigma|_{-1},\[\ssigma|_{-1}',\m|_2\]\]\bigg)\bigg].
\end{equation}
This is almost has the structure of \eq{cosusyalg}, but the momentum operator is missing. We therefore add and subtract~$2\P_1$  
\begin{equation}
\[\S,\S'\] = -2\P_1 + 2\P_1+\bigg[\,\M,\bigg( \[\s|_1,\ssigma|_{-1}'\] +\[\ssigma|_{-1},\s|_1'\]+\[\ssigma|_{-1},\[\ssigma|_{-1}',\m|_2\]\]\bigg)\bigg],
\end{equation}
and attempt to absorb $2\P_1$ into the commutator with $\M$. This can be achieved as follows. Since the gauge products are independent of the the position coordinate, we have the identity
\begin{equation}2\P_1 = \G^{-1}(2\P_1)\G, \end{equation}
which we can further write as
\begin{equation} 2\P_1 = \G^{-1}\[\Q,2\bm{\varpi}_1\] \G.\end{equation}
Moreover, since $\bm{\varpi}_1$ is a derivation of the star product we have
\begin{equation} 2\P_1 = \G^{-1}\[\Q+\m_2|_2,2\bm{\varpi}_1\]\G.\end{equation}
Absorbing the factors of $\G$ into the commutator gives
\begin{equation} 2\P_1 = \[\M,2\bm{\varpi}\],\end{equation}
where 
\begin{equation}\bm{\varpi}\equiv\G^{-1}\bm{\varpi}_1\G.\end{equation}
In this way we can absorb $2\P_1$ into the commutator with $\M$, giving 
\begin{equation}
\[\S,\S'\] = -2\P_1 +\bigg[\,\M,\bigg(2\bm{\varpi}+ \[\s|_1,\ssigma|_{-1}'\] +\[\ssigma|_{-1},\s|_1'\]+\[\ssigma|_{-1},\[\ssigma|_{-1}',\m|_2\]\]\bigg)\bigg].
\end{equation}
From this we can read off $\T$:
\begin{equation}
\T = 2\bm{\varpi}+ \[\s|_1,\ssigma|_{-1}'\] +\[\ssigma|_{-1},\s|_1'\]+\[\ssigma|_{-1},\[\ssigma|_{-1}',\m|_2\]\].\label{eq:T}
\end{equation}
In principle we could add an $[\M,\cdot]$-exact term, but we will show that this is not necessary. Note that $\T$ carries Ramond number zero.

Now we must confirm that $\T$ is in the small Hilbert space. For this we need the identities
\begin{eqnarray}
\[\n,\s|_1\] \lineup = \G^{-1}\[\n-\m_2|_0,\s_1|_1\] \G= \G^{-1}\[\s_1|_1,\m_2|_0\] \nonumber\\
\lineup = \[\s|_1,\m|_0\],\\
\[\n,\bm{\varpi}\] \lineup = \G^{-1}[\n-\m_2|_0,\bm{\varpi}_1]\G = \G^{-1}\p_1\G \nonumber\\
\lineup \equiv \p,\\
\[\n,\ssigma|_{-1}\] \lineup = \s|_{-1}.
\end{eqnarray}
Thus we find
\begin{eqnarray}
\[\n,\T\] \lineup = 2\p + \[\s|_1,\s|_{-1}'\] +\[\s|_{-1},\s|_1'\] \nonumber\\
\lineup\ \ \ +\[\[\s|_1,\m|_0\],\ssigma|_{-1}'\]-\[\ssigma|_{-1},\[\s|_1',\m|_0\]]+\[\s|_{-1},\[\ssigma|_{-1}',\m|_2\]\]-\[\ssigma|_{-1},\[\s|_{-1}',\m|_2\]\].
\end{eqnarray}
The first three terms cancel as follows: 
\begin{eqnarray}
2\p + \[\s|_1,\s|_{-1}'\] +\[\s|_{-1},\s|_1'\] \lineup = \G^{-1}\Big(2\p_1+\[\s_1|_1,\s_1|_{-1}'\]+\[\s|_{-1},\s|_1'\]\Big)\G  \nonumber\\
\lineup = \G^{-1}\Big(2\p_1+\[\s_1,\s_1'\]\Big)\G \nonumber\\
\lineup = 0,
\end{eqnarray}
where we used \eq{m1susyalg}. For the remaining terms, note that
\begin{equation}
\[\s|_{-1},\ssigma|_{-1}\]=0,
\end{equation}
since products cannot carry Ramond number $-2$. Therefore we can rearrange
\begin{eqnarray}
[\n,\T]\lineup=\[\[\s|_1,\m|_0\],\ssigma|_{-1}'\]-\[\ssigma|_{-1},\[\s|_1',\m|_0\]]+\[\ssigma|_{-1}',\[\s|_{-1},\m|_2\]\]-\[\ssigma|_{-1},\[\s|_{-1}',\m|_2\]\]
\nonumber\\
\lineup = \bigg[\ssigma|_{-1}',\Big(\[\s|_1,\m|_0\]+\[\s|_{-1},\m|_2\]\Big)\bigg]-\bigg[\ssigma|_{-1},\Big(\[\s|_1',\m|_0\]+\[\s|_{-1}',\m|_2\]\Big)\bigg].
\label{eq:Tsm1}\end{eqnarray}
Consider the object in parentheses in the first term:
\begin{eqnarray}
\[\s|_1,\m|_0\]+\[\s|_{-1},\m|_2\] \lineup = \G^{-1}\Big(\[\s_1|_1,\m_2|_0\]+\[\s_1|_{-1},\m_2|_2\]\Big)\G\nonumber\\
\lineup = \G^{-1}\[\s_1,\m_2\]\big|_1\G\nonumber\\
\lineup = 0,
\end{eqnarray}
which vanishes because $s_1$ is a derivation of the star product. The object in parentheses in the second term vanishes for the same reason after interchanging $\eps_{a}$ and $\eps_{a}'$. Therefore we have found $\T$ satisfying conditions (T.a)-(T.c).

\subsection{Cyclic Ramond Number Decomposition and Cyclicity}

We now demonstrate that $\T$ satisfies conditions (T.d) and (T.e) provided the dynamical products and supersymmetry products satisfy (M.d)-(M.e) and (S.d)-(S.e). 

The first step is to compute the cyclic Ramond number decomposition of $\T$. Since $\T$ carries Ramond number zero, it can have components at cyclic Ramond number zero and two:
\begin{equation}\T = \T|^0+\T|^2.\end{equation}
At first we might anticipate that $\T|^2$ will vanish, since it must vanish when operating on two Ramond states (since $\T$ has Ramond number 0) and by cyclicity it should then vanish when operating on one. The exception to this reasoning is if $\T|^2$ is composed entirely of  a 1-string product, which of course cannot be cyclically permuted to a product with two Ramond inputs. Let us see how this occurs. We compute $\T|^2$ using 
\begin{eqnarray}
\pi_1\T|^2\pi_{n+1} \lineup = \pi_1^1\T\pi_{n+1}^1\nonumber\\
\lineup = \pi_1^1\Big(2\bm{\varpi}+ \s|_1\ssigma|_{-1}' - \ssigma|_{-1}'\s|_1-\s|_1'\ssigma|_{-1} + \ssigma|_{-1}\s|_1' \nonumber\\
\lineup\ \ \ \ \ \ + \ssigma|_{-1}\ssigma|_{-1}'\m|_2 -\ssigma|_{-1}'\m|_2\ssigma|_{-1} + \ssigma|_{-1}'\m|_2\ssigma|_{-1} + \m|_2 \ssigma|_{-1}'\ssigma|_{-1}\Big)\pi_{n+1}^1,\label{eq:Tcyc1}
\end{eqnarray}
where in the second step we substituted \eq{T} and expanded the commutators. Now we substitute the formulas \eq{RGinv}, \eq{Rsig} and \eq{Rm2} for the Ramond outputs of $\G^{-1}$, $\ssigma|_{-1}$ and $\m|_2$:
\begin{eqnarray}
\pi_1\T|^2\pi_{n+1} \lineup = \pi_1^1\Bigg[\Big(2\bm{\varpi}_1\G - 2\Xi\m_2|_0\bm{\varpi}_1\G\Big)+\Big(\s_1|_1\G\ssigma|_{-1}'-\Xi\m_2|_0\s_1|_1\G\ssigma|_{-1}'\Big)
+\Big(-\Xi \s'_1|_{-1}\s_1|_1\G\Big) \nonumber\\
\lineup\ \ \ \ \ \ \ \ \ +\Big(-\s'_1|_1\G\ssigma|_{-1}\Big)+\Big(\Xi\m_2|_0\s'_1|_1\G\ssigma|_{-1}
+\Xi \s_1|_{-1}\s'_1|_1\G\Big) +\Big(\Xi\s_1|_{-1}\G\ssigma|_{-1}'\m|_2\Big)\nonumber\\
\lineup \ \ \ \ \ \ \ \ \ +\Big(-\Xi\s'_1|_{-1}\m_2|_2\G\ssigma|_{-1}\Big) + \Big(\Xi\s_1|_{-1}\m_2|_2\G\ssigma'|_{-1}\Big)\Bigg]\pi_{n+1}^1\nonumber\\
\lineup = \pi_1^1\Big(2\bm{\varpi}_1\G - 2\Xi\m_2|_0\bm{\varpi}_1\G -\Xi\m_2|_0\s_1|_1\G\ssigma|_{-1}'+\Xi\m_2|_0\s'_1|_1\G\ssigma|_{-1}
-\Xi \s'_1|_{-1}\s_1|_1\G \nonumber\\
\lineup\ \ \ \ \ \ \ \ \ 
+\Xi \s_1|_{-1}\s'_1|_1\G -\Xi\s'_1|_{-1}\m_2|_2\G\ssigma|_{-1} + \Xi\s_1|_{-1}\m_2|_2\G\ssigma'|_{-1}\Big)\pi_{n+1}^1.
\end{eqnarray}
The terms in parentheses in the first step correspond sequentially to the terms in \eq{Tcyc1}. In the second step we dropped some terms which vanish by Ramond number counting. Continuing, we can insert commutators in some terms as follows: 
\begin{eqnarray}
\pi_1\T|^2\pi_{n+1} \lineup =\pi_1^1\Big(2\bm{\varpi}_1\G + 2\Xi\bm{\varpi}_1\m_2|_0\G -\Xi\[\m_2|_0,\s_1|_1\]\G\ssigma|_{-1}'
-\Xi \[\s'_1|_{-1},\s_1|_1\]\G +\Xi\[\m_2|_0,\s'_1|_1\]\G\ssigma|_{-1}\nonumber\\
\lineup\ \ \ \ \ \ \ \ \ 
+\Xi \[\s_1|_{-1},\s'_1|_1\]\G -\Xi\[\s'_1|_{-1},\m_2|_2\]\G\ssigma|_{-1} + \Xi\[\s_1|_{-1},\m_2|_2\]\G\ssigma'|_{-1}\Big)\pi_{n+1}^1\nonumber\\
 \lineup =\pi_1^1\Big(2\bm{\varpi}_1\G + 2\Xi\bm{\varpi}_1\m_2|_0\G -\Xi[\m_2,\s_1]|_1\G\ssigma|_{-1}'+\Xi[\m_2,\s'_1]|_1\G\ssigma|_{-1}
+\Xi [\s_1,\s'_1]\G \Big)\pi_{n+1}^1.
\end{eqnarray}
Using \eq{m1susyalg} and the fact that $s_1$ is a derivation of the star product, we find
\begin{equation}
\pi_1\T|^2\pi_{n+1} =\pi_1^1\Big(2\bm{\varpi}_1\G + 2\Xi\bm{\varpi}_1\m_2|_0\G 
-2\Xi \p_1\G \Big)\pi_{n+1}^1.\label{eq:Tcyc2}
\end{equation}
To go further we will need to know something about how $\varpi_1$ commutes with $\Xi$. In appendix \ref{app:operator} we show
\begin{equation}
\Xi\varpi_1\Xi = 0.\label{eq:op1}
\end{equation}
Taking the commutator with $\eta$ implies the identity
\begin{equation}
[\Xi,\varpi_1] = \Xi p_1\Xi.
\end{equation}
Applying this to \eq{Tcyc2} gives 
\begin{eqnarray}
\pi_1\T|^2\pi_{n+1}\lineup =\pi_1^1\Big(2\bm{\varpi}_1\G - 2\bm{\varpi}_1\Xi\m_2|_0\G + 2\Xi \p_1\Xi\m_2|_0\G
-2\Xi \p_1\G \Big)\pi_{n+1}^1\nonumber\\
\lineup = \pi_1^1\Big(2\bm{\varpi}_1(1- \Xi\m_2|_0)\G -  2\Xi \p_1(1-\Xi\m_2|_0)\G\Big)\pi_{n+1}^1\nonumber\\
\lineup = \pi_1^1\Big(2\bm{\varpi}_1\G^{-1}\G -  2\Xi \p_1\G^{-1}\G\Big)\pi_{n+1}^1.
\end{eqnarray}
Canceling the $\G$s we find
\begin{equation}
\T|^2 = 2(\bm{\varpi}_1-\Xi\p_1)|^2.
\end{equation}
As anticipated, $\T|^2$ is composed entirely of a 1-string product. However, it is not obvious that this operator preserves the constraint on the Ramond string field in the restricted space. To address this question it is sufficient to consider the action of $\varpi_1-\Xi p_1$ on $\mathscr{X}$. In appendix \ref{app:operator} we will prove the identity
\begin{equation}
(\varpi_1 - \Xi p_1)\mathscr{X} = \frac{1}{\sqrt{2}}\psi_0 b_0\delta(\beta_0),\label{eq:op2}
\end{equation}
where $\psi_0$ is the zero mode of the worldsheet fermion
\begin{equation}
\psi_0 = \eps_{a} (C\Gamma^\mu)_{ab} \eps'_{b}\oint_{|z|=1}\frac{dz}{2\pi i}\frac{\psi_\mu(z)}{\sqrt{z}},
\end{equation}
and show that the operator \eq{op2} is preserved when acting $\mathscr{X}\mathscr{Y}$. Therefore $\T$ preserves the constraint on the Ramond string field, as required by condition (T.d). In appendix \ref{app:operator} we will also show that the operator \eq{op2} is BPZ odd, which will be important in a moment. 

We now turn to the proof of cyclicity. First we will consider the cyclic Ramond number zero component of~$\T$:
\begin{equation}
\langle \Omega|\pi_2\T|^0 = \langle \omega_S|\pi_2^0\T\ \ \ \mathrm{on}\ \ \Hr.
\end{equation}
As in subsection \ref{subsec:susycyc}, the argument goes more easily if we allow ourselves to contract with states in the large Hilbert space in intermediate steps. Therefore we will demonstrate the stronger relation 
\begin{equation}
\langle \omega_L|\pi_2^0\T=0,
\end{equation}
which holds for arbitrary states in $\mathcal{H}$. The computation is straightforward: 
\begin{eqnarray}
\langle \omega_L|\pi_2^0\T\lineup = \langle \omega_L|\pi_2^0\Big(2\bm{\varpi}+ \[\s|_1,\ssigma|_{-1}'\] +\[\ssigma|_{-1},\s|_1'\]+\[\ssigma|_{-1},\[\ssigma|_{-1}',\m|_2\]\]\Big)\nonumber\\
\lineup = \langle \omega_L|\pi_2^0\Big(2\bm{\varpi}+ \s|_1\ssigma|_{-1}'-\s|_1'\ssigma|_{-1}-\m|_2\ssigma|_{-1}'\ssigma|_{-1}\Big)\nonumber\\
\lineup = \langle \omega_L|\pi_2^0\Big(2\bm{\varpi}+ \s \ssigma|_{-1}'-\s'\ssigma|_{-1}-\m\ssigma|_{-1}'\ssigma|_{-1}\Big)\nonumber\\
\lineup = \langle \omega_L|\pi_2^0\Big(2\bm{\varpi}_1\G+ \s_1\G\ssigma|_{-1}'-\s_1'\G\ssigma|_{-1}-\m_2\G\ssigma|_{-1}'\ssigma|_{-1}\Big)\nonumber\\
\lineup = 0.
\end{eqnarray}
In the second step we noted that $\ssigma|_{-1}$ necessarily produces a Ramond output, and therefore vanishes against $\pi_2^0$. For the same reason, in the third step we drop the Ramond number restriction on $\s$ and $\m$. In the fourth step, we used that $\G$ is cyclic when it produces only NS outputs, and finally we obtain zero since $\varpi_1,s_1$ and $m_2$ are cyclic with respect to the large Hilbert space symplectic form. Next we consider cyclicity of the cyclic Ramond number 2 component of $\T$. For Ramond states $A$ and $B$ in the restricted space we have
\begin{eqnarray}
\Omega\Big(A,(\varpi_1-\Xi p_1)B\Big)\lineup = \omega_S\Big(\mathscr{Y}A,(\varpi_1-\Xi p_1)\mathscr{X}\mathscr{Y}B\Big)\nonumber\\
\lineup = \frac{1}{\sqrt{2}}\omega_S\Big(\mathscr{Y}A,\psi_0 b_0\delta(\beta_0)\mathscr{Y}B\Big)\nonumber\\
\lineup = -(-1)^{\deg(A)}\frac{1}{\sqrt{2}}\omega_S\Big(\psi_0 b_0\delta(\beta_0)\mathscr{Y}A,\mathscr{Y}B\Big).
\end{eqnarray}
In the last step we used the fact that $\psi_0 b_0\delta(\beta_0)$ is BPZ odd. Continuing
\begin{eqnarray}
\Omega\Big(A,(\varpi_1-\Xi p_1)B\Big)\lineup = -(-1)^{\deg(A)}\omega_S\Big((\varpi_1-\Xi p_1)\mathscr{X}\mathscr{Y}A,\mathscr{Y}B\Big)\nonumber\\
\lineup = -(-1)^{\deg(A)}\omega_S\Big((\varpi_1-\Xi p_1)A,\mathscr{Y}B\Big)\nonumber\\
\lineup = -(-1)^{\deg(A)}\Omega\Big((\varpi_1-\Xi p_1)A,B).
\end{eqnarray}
This completes the proof of cyclicity of $\T$. In summary, we have shown that the supersymmetry algebra can be expressed through \eq{cosusyalg}, with an explicit $\T$ satisfying all required properties (T.a)-(T.e).

\section{Supersymmetry and the $S$-matrix}
\label{sec:min}

It is interesting to illustrate how the supersymmetry transformation of string field theory is related to the usual on-shell supersymmetry which operates on open string scattering amplitudes. In string field theory we can derive the $S$-matrix in the standard way by gauge fixing and deriving Feynman rules. However, the theory of $A_\infty$ algebras gives an elegant but equivalent alternative via what is known as the {\it minimal model}. The minimal model is defined by a map (an $A_\infty$-quasi-isomorphism) which takes the $A_\infty$ algebra $\M$ into an $A_\infty$ algebra $\M_\mathrm{min}$ which operates on states satisfying the mass shell condition. The multi-string products of $\M_\mathrm{min}$ represent multi-string scattering amplitudes.

Let us review the definition of the minimal model. Since the construction is in principle well-known \cite{Kajiura,Kontsevich,Konopka}, we will mostly content ourselves with providing the formulas. See especially \cite{Konopka} for recent discussion in the context of superstring field theory, which motivates the construction from the perspective of homological perturbation theory. The first step is to  define a subspace of physical states $\mathcal{H}_p$ where we wish to define the minimal model. We require that the subspace contains all elements of the cohomology of $Q$,
\begin{equation}H^*(Q)\subseteq\mathcal{H}_p\subseteq\Hr,\end{equation}
and is closed under the action of the BRST operator. In the mathematics literature, it is usually assumed that $\mathcal{H}_p=H^*(Q)$, but this does not quite give the $S$-matrix as usually expressed by Feynman rules. We define a projection operator $\Pi$ which maps $\Hr$ into $\mathcal{H}_p$:
\begin{equation}\Pi: \Hr \to \mathcal{H}_p,\ \ \ \Pi^2 = \Pi.\end{equation}
Since $\mathcal{H}_p$ is closed under the action of $Q$, we have
\begin{equation}[Q,\Pi] = 0.\end{equation}
Note that $\mathbb{I}-\Pi$ projects onto a complimentary subspace where $Q$ contains no cohomology. Therefore $Q$ has a contracting homotopy operator $Q^+$ on this subspace:
\begin{equation}[Q,Q^+] = \mathbb{I}-\Pi.\label{eq:propagator}\end{equation}
$Q^+$ is degree odd, ghost number $-1$ and picture zero. In addition, we assume that $Q^+$ satisfies
\begin{eqnarray}(Q^+)^2 \lineup = 0, \label{eq:H2}\\
Q^+\Pi \lineup = \Pi Q^+ = 0. \label{eq:HPi}\end{eqnarray}
For string field theory amplitudes computed in Siegel gauge, the physical subspace $\mathcal{H}_p$ consists of states satisfying the mass shell condition $L_0=0$. The projector onto this subspace may be formally represented as
\begin{equation}\Pi = e^{-\infty L_0}.\end{equation}
The contracting homotopy operator $Q^+$ is precisely the Siegel gauge propagator:
\begin{equation}Q^+ = \frac{b_0}{L_0} = b_0\int_0^\infty dt\, e^{-tL_0}.\end{equation}
It is clear that the Siegel gauge propagator satisfies \eq{propagator} and \eq{H2}, whereas we assume that \eq{HPi} holds in a formal sense.

Next we promote $\Pi$ and $Q^+$ to natural operations on the tensor algebra. We lift $\Pi$ to a cohomomorphism $\bm{\hat{\Pi}}$ which acts on an $n$-string state simply as
\begin{equation}\bm{\hat{\Pi}}\pi_n = \underbrace{\Pi\otimes... \otimes\Pi}_{n\ \mathrm{times}} \pi_n.\end{equation}
Clearly $\bm{\hat{\Pi}}^2 = \bm{\hat{\Pi}}$, and 
\begin{equation}
\bm{\hat{\Pi}}:T\Hr \to T\mathcal{H}_p.
\end{equation}
Also, $[\Q,\bm{\hat{\Pi}}]=0$. The contracting homotopy $Q^+$ is lifted into an operator ${\bf Q^+}$:
\begin{equation}
{\bf Q^+}\pi_{n+1} = \sum_{k=0}^n \underbrace{\mathbb{I}\otimes ... \otimes\mathbb{I}}_{k\ \mathrm{times}}\otimes Q^+\otimes \underbrace{\Pi\otimes...\otimes\Pi}_{n-k\ \mathrm{times}} \, \pi_{n+1}.
\end{equation}
Note that $\Q^+$ is not quite a coderivation because inputs to the right of $Q^+$ above are projected by $\Pi$. Nevertheless the coproduct acts in a simple way:
\begin{equation}\triangle {\bf Q^+} = \Big(\mathbb{I}_{T\H}\otimes'{\bf Q^+}+{\bf Q^+}\otimes' \bm{\hat{\Pi}} \Big)\triangle.\end{equation}
The rationale for the definition of $\Q^+$ is the property 
\begin{equation}[\Q,{\bf Q^+}] = \mathbb{I}_{T\H} -\bm{\hat{\Pi}},\end{equation}
which can be viewed as a tensor algebra analogue of \eq{propagator}. We also have 
\begin{equation}({\bf Q^+})^2 = 0,\ \ \ \ \ {\bf Q^+}\bm{\hat{\Pi}} = \bm{\hat{\Pi}}{\bf Q^+} = 0,\end{equation}
corresponding to \eq{H2} and \eq{HPi}.

The minimal model for an $A_\infty$ algebra $\M$ can be expressed in the form 
\begin{equation}
\M_\mathrm{min} = {\bf \hat{P}}\M{\bf \hat{I}},\label{eq:minimal}
\end{equation}
where the cohomomorphisms ${\bf \hat{P}}$ and ${\bf \hat{I}}$ are called {\it projection} and {\it inclusion} maps, respectively. The projection map ${\bf \hat{P}}$ takes an element of $T\Hr$ into an appropriate element of $T\H_p$, while the inclusion map ${\bf \hat{I}}$ takes an element of $T\H_p$ into an appropriate element of $T\Hr$. They are given by the formulas
\begin{eqnarray}
{\bf \hat{P}}\lineup = \bm{\hat{\Pi}}\frac{1}{1+\delta\M\Q^+},\\
{\bf \hat{I}}\lineup =\frac{1}{1+\Q^+\delta\M}\bm{\hat{\Pi}},
\end{eqnarray}
where $\delta\M$ is the interacting part of the $A_\infty$ algebra $\M$:
\begin{equation}\delta\M\equiv\M - \Q.\end{equation}
It is also useful to introduce a nonlinear generalization of $\Q^+$: 
\begin{equation}
\M^+ = \Q^+\frac{1}{1+\delta\M\Q^+}.
\end{equation}
With some computation one can establish the following properties:
\begin{eqnarray}
{\bf \hat{P}}{\bf \hat{I}} \lineup = \bm{\hat{\Pi}},\ \ \ \ \ \ \ \ \ \ \ \ \ \ \ \ \ \ \ \ \, {\bf \hat{I}}{\bf \hat{P}} = \mathbb{I}_{T\H} - [\M,\M^+],\Vspace\\
({\bf \hat{P}}{\bf \hat{I}})^2\lineup  = {\bf \hat{P}}{\bf \hat{I}},\ \ \ \ \ \ \ \ \ \ \ \ \ \ \ \ ({\bf \hat{I}}{\bf \hat{P}})^2 = {\bf \hat{I}}{\bf \hat{P},}\Vspace\\
\triangle {\bf \hat{P}} \lineup = ({\bf \hat{P}}\otimes'{\bf \hat{P}})\triangle,\ \ \ \ \ \ \ \ \,\triangle {\bf \hat{I}} = ({\bf \hat{I}}\otimes'{\bf \hat{I}})\triangle.\Vspace
\end{eqnarray}
which imply
\begin{eqnarray}
[\M_\mathrm{min},\M_\mathrm{min}] \lineup = 0,\\
\triangle \M_\mathrm{min} \lineup = (\M_\mathrm{min}\otimes'\bm{\hat{\Pi}} + \bm{\hat{\Pi}}\otimes'\M_\mathrm{min})\triangle.
\end{eqnarray}
In particular $\M_\mathrm{min}$ is nilpotent, and the second relation implies that it acts as a coderivation on $T\H_p$. Therefore the products of $\M_\mathrm{min}$ define an $A_\infty$ algebra on the subspace of physical states. 

The $n+1$ string product of $\M_\mathrm{min}$ defines the color-ordered $n+2$ string scattering amplitude:
\begin{equation}
\mathcal{A}(\Phi_1,...,\Phi_{n+2}) = \Omega(\Phi_1,M_{\mathrm{min},{n+1}}(\Phi_1,...,\Phi_{n+2})),\ \ \ \ \Phi_i\in \H_p.\label{eq:minS}
\end{equation}
To see that this identification is plausible, note that after substituting the formulas for ${\bf\hat{P}}$ and ${\bf\hat{I}}$ we may express \eq{minimal} in the form
\begin{equation}
\M_\mathrm{min} = \bm{\hat{\Pi}}\left(\Q + \delta\M\frac{1}{1+\Q^+\delta\M}\right)\bm{\hat{\Pi}}.\label{eq:minex}
\end{equation}
From this we can compute (for example) the 4-point amplitude by extracting the 3-string product. For the Siegel gauge amplitude we obtain 
\begin{eqnarray}
M_{\mathrm{min},3} \pi_3\lineup = \pi_1\bm{\hat{\Pi}}\left(\Q + \delta\M\frac{1}{1+\Q^+\delta\M}\right)\bm{\hat{\Pi}}\pi_3\nonumber\\
\lineup = \Pi \pi_1\Big(\M_3 -\M_2\Q^+\M_2\Big)\bm{\hat{\Pi}}\pi_3\nonumber\\
\lineup = \Pi\left(M_3 - M_2\left(\frac{b_0}{L_0}\otimes\Pi + \mathbb{I}\otimes\frac{b_0}{L_0}\right)\Big(M_2\otimes\mathbb{I}+\mathbb{I}\otimes M_2\Big)\right)(\Pi\otimes\Pi\otimes\Pi)\pi_3\nonumber\\
\lineup = \Pi\left(M_3 - M_2\left(\frac{b_0}{L_0}M_2\otimes\mathbb{I} + \mathbb{I}\otimes\frac{b_0}{L_0}M_2\right)\right)(\Pi\otimes\Pi\otimes\Pi)\pi_3.
\end{eqnarray}
In the last step we formally assumed that $b_0/L_0$ annihilates $\Pi$. The first term gives the contribution from the quartic vertex, and the second and third terms give the contributions from a pair of cubic vertices connected by a propagator in the $s$ and $t$ channels. Let us evaluate the amplitude on a pair of Ramond states $R_1,R_2$ and a pair of NS states $N_1,N_2$ in $\H_p$: 
\begin{eqnarray}
\mathcal{A}(R_1,R_2,N_1,N_2)\lineup  = \Omega(R_1,M_3(R_2,N_1,N_2))-\Omega\left(R_1,M_2\left(\frac{b_0}{L_0}M_2(R_2,N_1),N_2\right)\right)\nonumber\\
\lineup\ \ \ \ \ \ \ \ \ \ \ \ \ \ \ \ \ \ \ \ \ \ \ \ \ \ \ \ \ \ \ \ \ \ 
-\Omega\left(R_1,M_2\left(R_2,\frac{b_0}{L_0}M_2(N_1,N_2)\right)\right)\nonumber\\
\lineup  = \Omega(R_1,\mathscr{X}m_3(R_2,N_1,N_2))-\Omega\left(R_1,\mathscr{X}m_2\left(\frac{b_0\mathscr{X}}{L_0}m_2(R_2,N_1),N_2\right)\right)
\nonumber\\
\lineup\ \ \ \ \ \ \ \ \ \ \ \ \ \ \ \ \ \ \ \ \ \ \ \ \ \ \ \ \ \ \ \ \ \ \ \,
-\Omega\left(R_1,\mathscr{X}m_2\left(R_2,\frac{b_0}{L_0}M_2(N_1,N_2)\right)\right)\nonumber\\
\lineup  = \omega_S(R_1,m_3(R_2,N_1,N_2))-\omega_S\left(R_1,m_2\left(\frac{b_0\mathscr{X}}{L_0}m_2(R_2,N_1),N_2\right)\right)
\nonumber\\
\lineup\ \ \ \ \ \ \ \ \ \ \ \ \ \ \ \ \ \ \ \ \ \ \ \ \ \ \ \ \ \ \ \ \ \ \ \,
-\omega_S\left(R_1,m_2\left(R_2,\frac{b_0}{L_0}M_2(N_1,N_2)\right)\right).
\end{eqnarray}
Here we simplified the amplitude knowing the form of the products with Ramond output and using the Ramond constraint $\mathscr{X}\mathscr{Y}=1$. Note that the diagram containing an intermediate Ramond state inherits a factor of $\mathscr{X}$ in the propagator, as would be expected from the propagator as derived by gauge fixing the Ramond kinetic term.

Now let us discuss supersymmetry. By analogy to \eq{minimal}, one might guess that the supersymmetry transformation in the minimal model will be described by
\begin{equation}\S_\mathrm{min} = {\bf\hat{P}}\S{\bf\hat{I}},\end{equation}
which acts as a coderivation on $T\H_p$. Assuming that $[S_1,Q^+] = [S_1,\Pi]=0$, which holds in Siegel gauge, this can be written explicitly as 
\begin{equation}
\S_\mathrm{min} = \bm{\hat{\Pi}}\left(\S_1 + \frac{1}{1+\delta\M\Q^+}\delta\S\frac{1}{1+\Q^+\delta\M}\right)\bm{\hat{\Pi}},\label{eq:Sminex}
\end{equation}
where 
\begin{equation}\delta\S \equiv \S -\S_1.\end{equation}
If $\S_\mathrm{min}$ is a symmetry of the minimal model, we expect
\begin{equation}[\S_\mathrm{min},\M_\mathrm{min}] = 0.\end{equation}
To see that this relation holds, compute
\begin{eqnarray}
[\S_\mathrm{min},\M_\mathrm{min}] \lineup ={\bf\hat{P}}\S{\bf\hat{I}}{\bf\hat{P}}\M{\bf\hat{I}}-{\bf\hat{P}}\M{\bf\hat{I}}{\bf\hat{P}}\S{\bf\hat{I}}\nonumber\\
\lineup ={\bf\hat{P}}\S(\mathbb{I}_{T\H} - [\M,\M^+])\M{\bf\hat{I}}-{\bf\hat{P}}\M(\mathbb{I}_{T\H} - [\M,\M^+])\S{\bf\hat{I}}\nonumber\\
\lineup ={\bf\hat{P}}\S\M(\M^+\M{\bf\hat{I}})-({\bf\hat{P}}\M\M^+)\M\S{\bf\hat{I}}.
\end{eqnarray}
Consider
\begin{eqnarray}
\M^+\M{\bf\hat{I}}\lineup = \Q^+\frac{1}{1+\delta\M\Q^+}\M\frac{1}{1+\Q^+\delta\M}\bm{\hat{\Pi}}\nonumber\\
\lineup = \frac{1}{1+\Q^+\delta\M}\Q^+(\Q+\delta\M)\frac{1}{1+\Q^+\delta\M}\bm{\hat{\Pi}}\nonumber\\
\lineup = \frac{1}{1+\Q^+\delta\M}([\Q, \Q^+] +\Q^+\delta\M)\frac{1}{1+\Q^+\delta\M}\bm{\hat{\Pi}}\nonumber\\
\lineup = \frac{1}{1+\Q^+\delta\M}(\mathbb{I}_{T\mathcal{H}} - \bm{\hat{\Pi}} +\Q^+\delta\M)\frac{1}{1+\Q^+\delta\M}\bm{\hat{\Pi}}\nonumber\\
\lineup = \frac{1}{1+\Q^+\delta\M}(\mathbb{I}_{T\mathcal{H}} +\Q^+\delta\M)\frac{1}{1+\Q^+\delta\M}\bm{\hat{\Pi}}-\frac{1}{1+\Q^+\delta\M}\bm{\hat{\Pi}}\nonumber\\
\lineup = \frac{1}{1+\Q^+\delta\M}\bm{\hat{\Pi}}-\frac{1}{1+\Q^+\delta\M}\bm{\hat{\Pi}}\nonumber\\
\lineup = 0.
\end{eqnarray}
Similarly one may show
\begin{equation}{\bf\hat{P}}\M\M^+ = 0.\end{equation}
Therefore $\S_\mathrm{min}$ is a symmetry of the minimal model.

To understand what this implies about scattering amplitudes, let us expand the commutator $[\S_\mathrm{min},\M_\mathrm{min}]$ using equations \eq{minex} and \eq{Sminex}:
\begin{eqnarray}
0 \lineup = \bm{\hat{\Pi}}\left(\left[\S_1,\delta\M\frac{1}{1-\Q^+\delta\M}\right] - \left[\Q, \frac{1}{1+\delta\M\Q^+}\delta\S\frac{1}{1+\Q^+\delta\M}\right]\right)\bm{\hat{\Pi}}\nonumber\\
\lineup\ \ \ + \bm{\hat{\Pi}}\frac{1}{1+\delta\M\Q^+}\delta\S\frac{1}{1+\Q^+\delta\M}\bm{\hat{\Pi}}\delta\M\frac{1}{1+\Q^+\delta\M}\bm{\hat{\Pi}}
-\bm{\hat{\Pi}}\delta\M\frac{1}{1+\Q^+\delta\M}\bm{\hat{\Pi}}\frac{1}{1+\delta\M\Q^+}\delta\S\frac{1}{1+\Q^+\delta\M}\bm{\hat{\Pi}}.\ \ \ \ \ \
\end{eqnarray}
Note that the second two terms contain the projector $\Pi$ between multi-string products. Such terms will only contribute if the external momenta are adjusted so as to produce intermediate states on the mass shell. Therefore for generic external momenta this relation simplifies to 
\begin{equation}
0 = \bm{\hat{\Pi}}\left(\left[\S_1,\delta\M\frac{1}{1-\Q^+\delta\M}\right] - \left[\Q, \frac{1}{1+\delta\M\Q^+}\delta\S\frac{1}{1+\Q^+\delta\M}\right]\right)\bm{\hat{\Pi}},\ \ \ \ \ \ \ \ (\mathrm{for\ generic\ momenta}).
\end{equation}
Moreover, for physical scattering amplitudes the external states will be BRST invariant. Therefore the second term, which is $\Q$ exact, will drop out. All that remains is the first term, which can be written 
\begin{equation}
[\S_1,\M_\mathrm{min}]=0,\ \ \ \ \ \ (\mathrm{for\ BRST\ invariant\ states\ at\ generic\ momenta}).
\end{equation}
All nonlinear terms in the supersymmetry transformation have dropped out. With the identification \eq{minS}, we conclude that scattering amplitudes satisfy 
\begin{equation}
\mathcal{A}(S_1\Phi_1,\Phi_2,...,\Phi_n) + \mathcal{A}(\Phi_1,S_1\Phi_2,...,\Phi_n)+...+ \mathcal{A}(\Phi_1,\Phi_2,...,S_1\Phi_n)=0,
\end{equation}
where $\Phi_i$ are BRST invariant states with generic momenta. This is the expected statement of supersymmetry at the level of the $S$-matrix.

\vspace{.5cm}

\noindent{\bf Acknowledgments}

\vspace{.25cm}

\noindent The author thanks S. Konopka, I. Sachs, Y. Okawa, and H. Kunitomo for discussions. This work is supported in part by the DFG Transregional Collaborative Research Centre TRR 33 and the DFG cluster of excellence Origin and Structure of the Universe.

\begin{appendix}

\section{Operator Identities}
\label{app:operator}

In this appendix we derive the equations \eq{op1} and \eq{op2} used in the computation of the supersymmetry algebra. We recall the bosonization formulas:
\begin{eqnarray}
\eta(z) \lineup = \d \Theta(\gamma(z)) = -\frac{\d}{\d z}\int \frac{dx}{x} e^{-x\gamma(z)},\\
\xi(x) \lineup = \Theta(\beta(z)) = -\int \frac{dx}{x} e^{-x\beta(z)},\label{eq:xibos}\\
e^{-\phi}(z)\lineup = \delta(\gamma(z)) = \int dx\, e^{-x\gamma(z)},\\
e^{\phi}(z)\lineup = \delta(\beta(z)) = \int dx \,e^{-x\beta(z)},\\
e^{\phi}\eta(z)\lineup = \gamma(z),\\
\d\xi e^{-\phi}(z)\lineup = \beta(z).
\end{eqnarray}
The integrals above are performed with respect to formal even variable $x$ and should be understood algebraically, analogous to the Berezin integral of an odd variable. See \cite{OO} and section 10 of \cite{revisited} for recent discussion. For our purposes, the most important properties of the algebraic integration concern the definition of the delta function 
\begin{equation}
\delta(y) = \int dx\, e^{-yx},\ \ \ \ \int dx\, \delta(x)f(x) = f(0),
\end{equation}
and the fact that the measure transforms with a Jacobian determinant {\it without} absolute value under changes of variables. This can lead to unexpected signs. For example
\begin{equation}\delta(-x) = -\delta(x),\end{equation}
which appears from the Jacobian after the change of variables $x\to-x$. For computations in the $\beta\gamma$ CFT it is generally sufficient that the algebraic integration can be performed on Gaussians multiplied by a polynomial of the even variables. However, for computations in the large Hilbert space it is necessary that integration can be performed on functions which have singularities when even variables vanish. Such singularities can be removed by factoring out the contribution from $\xi$ zero mode and reducing the $\eta,\xi,e^\phi$ correlator to a $\beta\gamma$ correlator. The manner in which this is done is somewhat arbitrary, but one prescription is described in~\cite{RWaction}. 

Let us give a sample computation which will be useful in a moment. Consider the operator
\begin{equation}\xi e^{-\phi}(z) = \Theta(\beta(z))\delta(\gamma(z)),\end{equation}
which can be represented in terms of an algebraic integral,
\begin{equation}
\xi e^{-\phi}(z) = - \lim_{z'\to z}\int dx_1dx_2\frac{1}{x_1}e^{-x_1\beta(z')}e^{-x_2\gamma(z)}.
\end{equation}
To understand the limit $z\to z'$ it is necessary to normal order the exponentials: 
\begin{equation}
e^{-x_1\beta(z')}e^{-x_2\gamma(z)} = \exp\left(-\frac{x_1x_2}{z'-z}\right):e^{-x_1\beta(z')}e^{-x_2\gamma(z)}:\,.
\end{equation}
The singularity from the OPE can be absorbed into a change of variables in the algebraic integral $\frac{x_1}{z'-z}\to x_1$. Then we can take the limit $z\to z'$ to obtain 
\begin{eqnarray}
\xi e^{-\phi}(z) \lineup = - \int dx_1dx_2\frac{1}{x_1}e^{-x_1x_2} e^{-x_2\gamma(z)}\nonumber\\
\lineup = - \int dx_1dx_2\frac{1}{x_1} e^{-x_2(\gamma(z)+x_1)}.
\end{eqnarray}
We can now easily integrate over $x_2$ to find 
\begin{eqnarray}
\xi e^{-\phi}(z) \lineup =  - \int dx_1 \frac{1}{x_1}\delta(\gamma(z) +x_1).\nonumber\\
\end{eqnarray}
Performing the integral over $x_1$ gives
\begin{equation}\xi e^{-\phi}(z) = \Theta(\beta(z))\delta(\gamma(z)) = \gamma(z)^{-1}.\end{equation}
To see that this result is logical, note the OPE 
\begin{equation}
\xi e^{-\phi}(z)\gamma(0) = 1+\mathcal{O}(z),
\end{equation}
which naturally suggests the identification $\xi e^{-\phi}(z) = \gamma(z)^{-1}$.

Let us recall the formula for $\Xi$ given in \cite{RWaction}:
\begin{equation}
\Xi = \xi  + (\Theta(\beta_0)\eta\xi - \xi)P_{-3/2} + (\xi\eta\Theta(\beta_0) - \xi)P_{-1/2}.
\end{equation}
Here $P_n$ is the projector onto picture $n$. The operator $\xi$ is defined 
\begin{equation}\xi \equiv \oint_{|z|=1} \frac{dz}{2\pi i} f(z)\xi(z),\end{equation}
for some function $f(z)$ which is holomorphic in the vicinity of the unit circle and normalized so that $[\eta,\xi] = 1$. Finally we have the operator 
\begin{equation}\Theta(\beta_0) \equiv -\int \frac{dx}{x}e^{-x\beta_0},\end{equation}
defined by an algebraic integral analogously to \eq{xibos}.

\subsection{Proof of \eq{op2}}

Let us demonstrate the identity \eq{op2}
\begin{equation}(\varpi_1 - \Xi p_1) \mathscr{X} = \frac{1}{\sqrt{2}} \psi_0 b_0 \delta(\beta_0),
\end{equation}
which holds when acting on Ramond states the small Hilbert space at picture $-3/2$. Focus on the term $\Xi p_1\mathscr{X}$. In the small Hilbert space at picture $-3/2$ we can replace $\Xi$ with $\Theta(\beta_0)$. Inside the contour integral defining $p_1$ is the operator $e^{-\phi}(z) = \delta(\gamma(z))$, so we start by computing 
\begin{equation}\Theta(\beta_0)\delta(\gamma(z))\mathscr{X} = -\Theta(\beta_0)\delta(\gamma(z))\delta(\beta_0)G_0 +\Theta(\beta_0)\delta(\gamma(z))\delta'(\beta_0)b_0,
\end{equation}
where we substituted $\mathscr{X}$ in the form
\begin{equation}\mathscr{X} = -\delta(\beta_0)G_0+\delta'(\beta_0)b_0.
\end{equation}
Now compute 
\begin{equation}
\Theta(\beta_0)\delta(\gamma(z))\delta(\beta_0) = -\int dx_1dx_2 \frac{1}{x_1}e^{-x_1\beta_0}e^{-x_2\gamma(z)}\delta(\beta_0),
\end{equation}
Note that because 
\begin{equation}[\beta_0,\gamma(z)]=-\sqrt{z},\end{equation}
with $|z|=1$, the exponential of $\beta_0$ acts as a translation operator on the exponential of $\gamma(z)$, and we obtain
\begin{eqnarray}
\Theta(\beta_0)\delta(\gamma(z))\delta(\beta_0)
\lineup = -\int dx_1dx_2 \frac{1}{x_1}e^{-x_1\beta_0}e^{-x_2\gamma(z)}e^{x_1\beta_0}\delta(\beta_0)\nonumber\\
\lineup = -\int dx_1dx_2 \frac{1}{x_1}e^{-x_2(\gamma(z)+\sqrt{z}x_1)}\delta(\beta_0).
\end{eqnarray}
Making a substitution $\sqrt{z}x_1\to x_1$ and integrating over $x_2$ we find
\begin{eqnarray}
\Theta(\beta_0)\delta(\gamma(z))\delta(\beta_0)
\lineup = -\int dx_1 \frac{1}{x_1}\delta(\gamma(z) + x_1) \delta(\beta_0)\nonumber\\
\lineup = \gamma(z)^{-1}\delta(\beta_0).
\end{eqnarray}
Similarly we may compute
\begin{eqnarray}
\Theta(\beta_0)\delta(\gamma(z))\delta'(\beta_0)
\lineup = \frac{1}{\sqrt{z}}\Theta(\beta_0)\delta(\gamma(z))[\gamma(z),\delta(\beta_0)]\nonumber\\
\lineup = -\frac{1}{\sqrt{z}}\Theta(\beta_0)\delta(\gamma(z))\delta(\beta_0)\gamma(z)\nonumber\\
\lineup =  -\frac{1}{\sqrt{z}}\gamma(z)^{-1}\delta(\beta_0)\gamma(z)\nonumber\\
\lineup =  -\frac{1}{\sqrt{z}}\delta(\beta_0)+\frac{1}{\sqrt{z}}\gamma(z)^{-1}[\gamma(z),\delta(\beta_0)]\nonumber\\
\lineup =  -\frac{1}{\sqrt{z}}\delta(\beta_0)+\gamma(z)^{-1}\delta'(\beta_0).
\end{eqnarray}
Plugging in, we therefore find that 
\begin{equation}\Theta(\beta_0)\delta(\gamma(z))\mathscr{X} = \gamma(z)^{-1}\mathscr{X} +\frac{1}{\sqrt{z}}b_0\delta(\beta_0).
\end{equation}
Multiplying this equation with 
\begin{equation}\frac{1}{\sqrt{2}}\eps_{a} (C\Gamma^\mu)_{ab} \eps'_{b} \psi_\mu(z)\end{equation}
and integrating $z$ around the unit circle reproduces \eq{op2}.

Let us now check that \eq{op2} is consistent with the constraint $\mathscr{X}\mathscr{Y}=1$ on the Ramond string field. Note the relation
\begin{equation}
\delta(\beta_0)\delta(\gamma_0)\delta(\beta_0) = \delta(\beta_0),
\end{equation}
and associated identities
\begin{eqnarray}
\delta(\beta_0)\delta'(\gamma_0)\delta(\beta_0)  \lineup = \delta(\beta_0)[\delta(\gamma_0),\beta_0]\delta(\beta_0)\nonumber\\
\lineup = 0,\nonumber\\
\delta'(\beta_0)\delta'(\gamma_0)\delta(\beta_0)\lineup = [\gamma_0,\delta(\beta_0)]\delta'(\gamma_0)\delta(\beta_0)\nonumber\\
\lineup = -\delta(\beta_0)\gamma_0\delta'(\gamma_0)\delta(\beta_0)\nonumber\\
\lineup = -\delta(\beta_0)\Big([\beta_0,\gamma_0\delta(\gamma_0)] - \delta(\gamma_0)\Big)\delta(\beta_0)\nonumber\\
\lineup = \delta(\beta_0)\delta(\gamma_0)\delta(\beta_0) \nonumber\\
\lineup = \delta(\beta_0).
\end{eqnarray}
With this we can compute the action of $\mathscr{X}\mathscr{Y}$ on the right hand side of \eq{op2}:
\begin{eqnarray}
\mathscr{X}\mathscr{Y}\big(\psi_0 b_0\delta(\beta_0)\big)\lineup =(G_0 \delta(\beta_0) + b_0\delta'(\beta_0))(-c_0\delta'(\gamma_0))b_0\delta(\beta_0)\psi_0
\nonumber\\
\lineup = b_0 c_0 b_0 \delta'(\beta_0)\delta'(\gamma_0)\delta(\beta_0)\psi_0\nonumber\\
\lineup = \psi_0  b_0 \delta(\beta_0).
\end{eqnarray}
Therefore the operator \eq{op2} produces a Ramond string field in the restricted space. 

Next we need to show that the operator $\psi_0 b_0 \delta(\beta_0)$ is BPZ odd. If $\star$ denotes BPZ conjugation, we have
\begin{equation}
\Big(\psi_0 b_0 \delta(\beta_0)\Big)^\star = -  \delta(\beta_0^\star)b_0^\star \psi_0^\star,
\end{equation}
where the sign comes from reversing the order of the operators. If $\phi(z)$ is a primary of weight $h$, the modes $\phi_n$ BPZ conjugate as
\begin{equation}\phi_n^\star = e^{i\pi(n+h)}\phi_{-n}.\end{equation}
Therefore we have 
\begin{eqnarray}
\Big(\psi_0 b_0 \delta(\beta_0)\Big)^\star \lineup = - \delta\big(e^{3\pi i/2}\beta_0\big)\big(e^{2\pi i}b_0\big)\big( e^{i\pi/2} \psi_0\big)\nonumber\\
\lineup = - e^{-3\pi i/2 + 2\pi i +i \pi/2}\delta(\beta_0)b_0 \psi_0\nonumber\\
\lineup = \delta(\beta_0)b_0 \psi_0\nonumber\\
\lineup = - \psi_0 b_0\delta(\beta_0).
\end{eqnarray}
So the operator is BPZ odd as required for cyclicity.

\subsection{Proof of \eq{op1}}

Next we prove the identity \eq{op1}:
\begin{equation}\Xi \varpi_1\Xi = 0.\end{equation}
This identity holds trivially on all pictures except $-3/2$, where it is sufficient to demonstrate the relation
\begin{equation}
\Theta(\beta_0)\gamma(z)^{-1}\Theta(\beta_0) = 0
\end{equation}
for $|z|=1$. Substituting $\gamma(z)^{-1} = \Theta(\beta(z))\delta(\gamma(z))$ we have
\begin{equation}
\Theta(\beta_0)\gamma(z)^{-1}\Theta(\beta_0) = -\Theta(\beta(z)) \Big(\Theta(\beta_0)\delta(\gamma(z))\Theta(\beta_0)\Big).
\end{equation}
Focus on the expression in parentheses. Using algebraic integration we may write
\begin{eqnarray}
\Theta(\beta_0)\delta(\gamma(z))\Theta(\beta_0) \lineup = \int dx_1 dx_2 dx_3 \frac{1}{x_1x_3} e^{-x_1\beta_0} e^{-x_2\gamma(z)} e^{-x_3 \beta_0}\nonumber\\
\lineup = \int dx_1 dx_2 dx_3 \frac{1}{x_1x_3} e^{-x_2(\gamma(z)+\sqrt{z}x_1)} e^{-(x_1+x_3) \beta_0}\nonumber\\
\lineup = \int dx_1 dx_3 \frac{1}{x_1x_3} \delta(\gamma(z)+\sqrt{z}x_1) e^{-(x_1+x_3) \beta_0}.\nonumber\\
\end{eqnarray}
Next make a change of variables $x_1+x_3\to x_3$:
\begin{eqnarray}
\Theta(\beta_0)\delta(\gamma(z))\Theta(\beta_0)
\lineup = \int dx_1 dx_3 \frac{1}{x_1(x_3-x_1)} \delta(\gamma(z)+\sqrt{z}x_1) e^{-x_3 \beta_0}\nonumber\\
\lineup = \int dx_3 \frac{1}{(-\gamma(z)/\sqrt{z})(x_3+\gamma(z)/\sqrt{z})}e^{-x_3 \beta_0}\nonumber\\
\lineup = -z \int dx_3 \gamma(z)^{-1}e^{-x_3\beta_0}\gamma(z)^{-1}\nonumber\\
\lineup = - z\gamma(z)^{-1}\delta(\beta_0)\gamma(z)^{-1}.
\end{eqnarray}
With this we substitute $\gamma(z)^{-1} = \Theta(\beta(z))\delta(\gamma(z))$ to obtain
\begin{equation}
\Theta(\beta_0)\delta(\gamma(z))\Theta(\beta_0) = z \delta(\gamma(z)) \Theta(\beta(z))\delta(\beta_0)\Theta(\beta(z))\delta(\gamma(z)) = 0,
\end{equation}
which vanishes since $\Theta(\beta(z))^2=\xi(z)^2 = 0$. Multiplying by 
\begin{equation}-\frac{1}{\sqrt{2}}\eps_{a} (C\Gamma^\mu)_{ab} \eps'_{b}\Theta(\beta(z))\end{equation}
and integrating $z$ over the unit circle establishes \eq{op1}.

\end{appendix}

\end{document}